\documentclass[sigconf]{acmart} 
\AtBeginDocument{%
  \providecommand\BibTeX{{%
    \normalfont B\kern-0.5em{\scshape i\kern-0.25em b}\kern-0.8em\TeX}}}
    
\usepackage{acmart-taps}
\usepackage{xcolor}    
\usepackage{subcaption}
\usepackage{color,soul}
\usepackage{amsmath}

\newcommand{\todo}[1]{\textcolor{black}{#1}}
\newcommand{\smnote}[1]{\textcolor{black}{#1}}
\newcommand{\smchange}[1]{\textcolor{black}{#1}}
\newcommand{\rgchange}[1]{\textcolor{black}}

\newcommand{\rs}{reward-shaping}
\newcommand{\studyone}{\textit{hybrid} agent}

\newcommand{\studythree}{\emph{\rs} agent}

\newcommand{\rulesep}{\unskip\ \vrule\ }
\newcommand{\numtotalhntt}{6}

\definecolor{bluehl}{rgb}{0.8706,.921568,.9686274} 
\definecolor{purplehl}{rgb}{0.88627,0.815686,0.941176} 
\definecolor{orangehl}{rgb}{0.9843,0.898,0.83922} 
\DeclareRobustCommand{\hlblue}[1]{{\sethlcolor{bluehl}\hl{#1}}}
\newcommand{\ptone}[1]{\textcolor{black}{#1}}
\newcommand{\pttwo}[1]{\textcolor{black}{#1}}
\newcommand{\ptthree}[1]{\textcolor{black}{#1}}
\newcommand{\ptfour}[1]{\textcolor{black}{#1}}

\usepackage[strict]{changepage}
\usepackage{framed}
\definecolor{formalshade}{rgb}{0.95,0.95,1}
\newenvironment{formal}{%
  \MakeFramed{\advance\hsize-\width\FrameRestore}%
  \noindent\hspace{-4.55pt}
  \begin{adjustwidth}{}{7pt}%
  \vspace{0.5pt}\vspace{0.5pt}%
}
{%
  \vspace{0.5pt}\end{adjustwidth}\endMakeFramed%
}

\definecolor{shadecolor}{rgb}{0.95,0.95,1}

\copyrightyear{2023}
\acmYear{2023}
\setcopyright{rightsretained}
\acmConference[CHI '23]{Proceedings of the 2023 CHI Conference on Human Factors in Computing Systems}{April 23--28, 2023}{Hamburg, Germany}
\acmBooktitle{Proceedings of the 2023 CHI Conference on Human Factors in Computing Systems (CHI '23), April 23--28, 2023, Hamburg, Germany}\acmDOI{10.1145/3544548.3581348}
\acmISBN{978-1-4503-9421-5/23/04}

\begin{document}

\title{Navigates Like Me: Understanding How People Evaluate \\ Human-Like AI in Video Games}

\author{Stephanie Milani}
\email{smilani@andrew.cmu.edu}
\orcid{0000-0003-1150-4418}
\affiliation{
  \institution{Carnegie Mellon University}
  \streetaddress{5000 Forbes Avenue}
  \city{Pittsburgh}
  \state{Pennsylvania}
  \country{USA}
  \postcode{15213}
}
\author{Arthur Juliani}
\affiliation{%
  \institution{Microsoft Research}
  \streetaddress{}
  \city{New York}
  \state{New York}
  \country{USA}
  \postcode{}
}

\author{Ida Momennejad}
\affiliation{%
  \institution{Microsoft Research}
  \streetaddress{}
  \city{New York}
  \state{New York}
  \country{USA}
  \postcode{}
}

\author{Raluca Georgescu}
\affiliation{%
  \institution{Microsoft Research}
  \streetaddress{}
  \city{Cambridge}
  \state{}
  \country{United Kingdom}
  \postcode{}
}

\author{Jaroslaw Rzpecki}
\affiliation{%
 \institution{Monumo}
 \streetaddress{}
 \city{Cambridge}
 \state{}
 \country{United Kingdom}}

\author{Alison Shaw}
\affiliation{%
  \institution{Ninja Theory}
  \streetaddress{}
  \city{Cambridge}
  \state{}
  \country{United Kingdom}}

\author{Gavin Costello}
\affiliation{%
  \institution{Ninja Theory}
  \streetaddress{}
  \city{Cambridge}
  \state{}
  \country{United Kingdom}}

\author{Fei Fang}
\affiliation{
  \institution{Carnegie Mellon University}
  \streetaddress{5000 Forbes Avenue}
  \city{Pittsburgh}
  \state{Pennsylvania}
  \country{USA}
  \postcode{15213}
}

\author{Sam Devlin}
\affiliation{%
  \institution{Microsoft Research}
  \streetaddress{}
  \city{Cambridge}
  \state{}
  \country{United Kingdom}
  \postcode{}
}

\author{Katja Hofmann}
\affiliation{%
  \institution{Microsoft Research}
  \streetaddress{}
  \city{Cambridge}
  \state{}
  \country{United Kingdom}
  \postcode{}
}

\renewcommand{\shortauthors}{Milani, et al.}

\begin{abstract}
We aim to understand how people assess human likeness in navigation produced by people and artificially intelligent (AI) agents in a video game.
To this end, we propose a novel AI agent with the goal of generating more human-like behavior.
We collect hundreds of crowd-sourced assessments comparing the human-likeness of navigation behavior generated by our agent and baseline AI agents with human-generated behavior.
Our proposed agent passes a Turing Test, while the baseline agents do not. 
By passing a Turing Test, we mean that human judges could not quantitatively distinguish between videos of a person and an AI agent navigating. 
To understand what people believe constitutes human-like navigation, we extensively analyze the justifications of these assessments. 
This work provides insights into the characteristics that people consider human-like in the context of goal-directed video game navigation, which is a key step for further improving human interactions with AI agents. 
\end{abstract}




\maketitle

\section{Introduction}
Games are considered one of the oldest forms of human social interaction~\cite{tylor1879history,gamingtokens2013}. 
Throughout history, people have played games as important cultural and social bonding events~\cite{lavega2004traditional}, as teaching and learning tools~\cite{de2018games,lamsa2018games}, and for enjoyment~\cite{koster2013theory}. 
Today, video games have emerged as a popular form of structured play, inviting players to immerse themselves in captivating virtual worlds.
This immersion is vital to making these games enjoyable~\cite{christou2014interplay,michailidis2018flow}.

To enhance this immersive experience, game designers focus on creating believable non-player characters (NPCs) that can interact with players in diverse ways. 
A crucial part of believability is \textit{human-likeness} --- that is, the ability of the NPC to behave as a person would.
Because many player-NPC interactions are critical to the game, it is important that the NPCs behave believably to maintain immersion~\cite{conroy2011modeling,warpefelt2013mind,johansson2013non}.
Indeed, video game players find playing against more human-like agents more enjoyable~\cite{soni2008bots}.
Traditionally, game developers have designed NPCs to follow a predetermined set of actions. 
However, this approach can be both time-consuming and challenging, which has motivated designers to turn to artificial intelligence (AI) for assistance with NPC design. 
Due to this shift, there is a need for research into understanding what people perceive as human-like in AI agents.

At the same time, AI researchers have identified achieving complex human-like behavior as a critical milestone~\citep{brooks1998alternative,sloman1999sort,gil201920} towards developing agents that can flexibly collaborate with people in shared human-AI environments~\cite{carroll2019utility,wang2020human,guttman2021play} and various robotics applications~\citep{scheutz2007first}.
This goal is not satisfied by agents demonstrating a high proficiency level at the assigned task. 
For example, AI-powered vehicles must behave sufficiently human-like for human drivers to interpret, anticipate, and act in their presence~\citep{hecker2020learning}. 
As a result, understanding the behaviors that contribute to people's perceptions of human likeness is a foundational first step towards achieving general human-like behavior of artificial agents.

In this work, we contribute to the objective of developing human-like agents by identifying and understanding what constitutes human-like behaviors in a video game.
To scope our study, we focus on a 3D video game where agents must navigate from one point to another.
This form of navigation is pervasive in many video games, making it a key area of interest for game developers~\cite{alonso2020deep,dubois2021visualizing}: in embodied games, players must move from place to place to accomplish their goals or explore the world.
More generally, it is considered fundamental to embodied biological intelligence~\cite{o1978hippocampus,hafting2005microstructure}, making it of interest to cognitive scientists~\cite{richardson2011video,murias2016effects} and researchers interested in intelligent behavior~\cite{wijmans2023emergence}.
It has also been a key area of interest in HCI~\cite{vainio2010review} due to how people (or robots) navigate in real, augmented, or entirely virtual spaces.

To study navigation in video games, we leverage the recently-proposed Human Navigation Turing Test (HNTT)~\citep{devlin2021navigation}, in which human judges indicate which of two videos demonstrates more human-like behavior. 
The judges then justify their decision and indicate their certainty about their choice. 
In that work, the authors compared the accuracy of the human-likeness assessments to random chance but did not instantiate a statistical test to definitively conclude whether an agent passed the HNTT.
According to their assessment, both studied AI agents did not pass the HNTT.
As a result, producing an agent that passes the HNTT is still an open challenge.
To this end, we design a novel agent to pass the HNTT.
To assist with our design of a human-like agent, we inspect the resulting behavior of the two baseline agents from prior work~\cite{devlin2021navigation}. 
With these insights, we design our novel agent --- the \emph{reward-shaping} agent --- using simple and intuitive techniques. 

We then conduct a behavioral study on Amazon Mechanical Turk (MTurk) of the HNTT to investigate the behavior of our agent and the baselines. 
To determine whether agents pass the HNTT, we propose a firm criterion: a statistical test that determines whether human judges distinguish between human and agent behavior at a level that is \textit{significantly different} from chance.
We then validate the conclusion of previous work: the two baseline agents are not sufficiently human-like because they do not pass the HNTT.
In contrast, human judges cannot reliably distinguish between the behavior of our \emph{reward-shaping} agent from one controlled by a person.
To our knowledge, this agent is the first to pass the HNTT.

To understand these assessments, we analyze the free-form responses to determine which characteristics people believe are representative of human and AI navigation behavior.
We annotate the responses with codes that summarize the provided rationale. 
Using these annotations, we find that there are key differences between how people characterize human-like and non-human-like behavior. 
Specifically, we find that people utilize the same high-level characteristics when describing human-like and non-human-like behavior, but the presence or absence of these characteristics strongly informs their judgments.

Based on the findings of our analysis, we summarize considerations when developing and evaluating the human likeness of AI agents.
For example, considering the end use of the agent is critical for defining what is meant by human like and designing a study accordingly.
In summary, we make the following contributions.
\begin{enumerate}
    \item We contribute a novel \emph{\rs{}} agent that exhibits more human-like navigation behavior.
    \item We conduct a behavioral study to assess: a) whether people reliably distinguish the behavior produced by the AI agents from that generated by people and b) what characteristics people believe are indicative of human-like behavior.
    \item We conduct an extensive analysis of the resulting data.
    We propose a firm criterion to determine whether an agent passes the HNTT and find that only our \emph{\rs{}} agent passes the HNTT according to this metric.
    We analyze the free-form responses to determine the characteristics that people believe are representative of human-like behavior.
    \item Based on our findings, we propose concrete suggestions for developing and evaluating human-like AI.
\end{enumerate}

\section{Related Work}
\label{sec:relwork}

Researchers have taken various approaches to address the challenge of developing believable AI agents in games, including learning from demonstrations~\cite{karpov2013believable,mendoncca2015simulating,jacob2022modeling}, reinforcement learning~\cite{arzate2018hrlb,fujii2013evaluating,zhao2019multi,mendoncca2015simulating}, and more~\cite{thurau2004learning,miranda2016neuroevolution}.
We focus on \textit{reinforcement learning}~\citep{sutton2018reinforcement} because it provides a generally-applicable set of algorithms for learning to control agents in settings including (but not limited to) modern game environments~\citep{szita2012reinforcement,lample2017playing,guss2019neurips,vinyals2019grandmaster,alonso2020deep}. 
It also offers significant benefits as an approach for generating navigation behavior~\cite{alonso2020deep}. 
In particular, the use of reinforcement learning may enable more complex navigation abilities (such as grappling or teleportation) and alleviate game designers from the labor-intensive procedure of the most popular alternative method to produce this behavior~\cite{mcanlis2008intrinsic}.

In reinforcement learning, an agent learns to accomplish a task by maximizing a reward, or score, that tells the agent how well it is performing. 
Although agents learn effective navigation by maximizing this reward, they make no consideration for the \textit{style} with which they act~\citep{alonso2020deep}.
If these approaches are to be adopted in commercial game development, practitioners have firmly asserted that controlling style is essential~\citep{jacob2020s}. 
As an extreme example, reinforcement learning approaches that have recently defeated world champion human players at modern games demonstrated unusual behaviors~\cite{holcomb2018overview} that made collaborative play between human and AI in mixed teams far less successful~\citep{berner2019dota}.
Simply maximizing the task-specific reward signal is unlikely to produce human-like agents.

\begin{figure*}[t]
\centering 
\includegraphics[width=.95\textwidth]{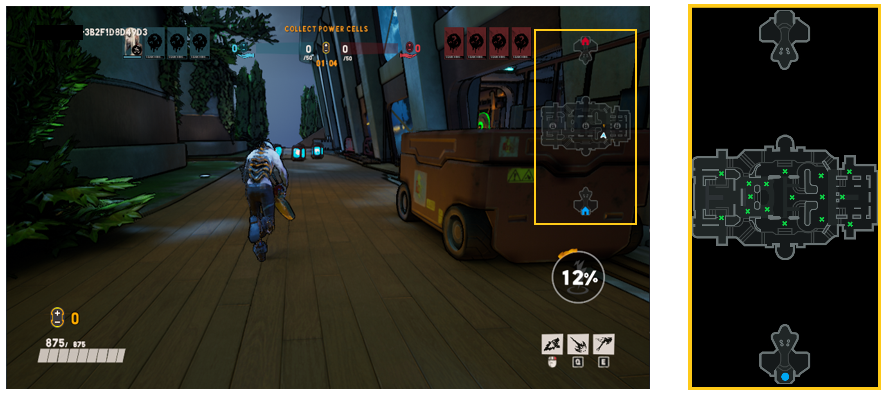}
\caption{\textbf{Navigation task as observed by study participants (screenshot, left), and detail of the mini map of the game level (right).} 
Agents spawn on the island outside of the main map, which is shown in the bottom portion of the mini map on the right. 
They must jump to the main area and navigate to the goal location.
The light blue containers in the left screenshot represent the goal location.}
\label{fig:be-screenshot}
\Description{Screenshot of navigation task as observed by study participants and screenshot of minimap of the game level. The agent is viewed in third person and is running toward the horizon.}
\end{figure*}

\pttwo{Reward shaping~\cite{ng1999policy,Wiewiora2010} is a simple yet powerful technique that allows practitioners to clearly specify the desired agent behavior.
This approach involves crafting a reward signal that provides dense feedback to the agent.
It is an intuitive way for those without a machine learning background to control the agent's behavior by specifying objectives instead of dedicating time to optimizing unintuitive hyperparameters.
Additionally, reward shaping can be used with any reinforcement learning algorithm, making it possible to swap in and out the underlying algorithm as needed.
We utilize reward shaping to generate more human-like behavior.}

There is no standard set of metrics for evaluating human-like AI.
One paradigm involves measuring human similarity with proxy metrics for human judgments.
Some work measures the task performance of the AI~\cite{zhao2019multi,thurau2004learning}, but this metric is an insufficient proxy for human similarity.
Other work assesses how well the AI agent can predict the following human action~\cite{jacob2022modeling} or align its behavior with people~\cite{tence2010automatable,miranda2016neuroevolution}, but these metrics do not include actual human evaluations.
They do not assess whether people can accurately distinguish the AI player from the human one, which is vital for assessing human likeness in games~\cite{glende2004agent} and beyond~\cite{zador2022toward}.

Studies with human evaluations tend to be small-scale surveys to understand the opinions regarding human-likeness~\cite{fujii2013evaluating,mendoncca2015simulating}.
They often offer only a preliminary investigation into the specific characteristics that inform these beliefs and typically do not include a form of Turing test~\citep{kim2018performance}, a well-established framework for addressing these problems~\cite{glende2004agent}.
Work that uses a Turing test often does not investigate the behaviors or provide concrete metrics~\cite{mozgovoy2011behavior}, or it focuses on assessing the full spectrum of game behaviors~\cite{mendoncca2015simulating,fujii2013evaluating,arzate2018hrlb}. 
Due to the complexity of these games and the resulting behaviors, providing concrete recommendations to game designers is challenging.
In contrast, we focus on a specific but widely-used behavior: point-to-point navigation.
To perform our assessment, we utilize the setup of the recently-proposed Human Navigation Turing Test \citep{devlin2021navigation}; however, we propose and perform a deeper evaluation of human assessments of AI and human behavior.

\section{Background and Preliminaries}
\label{sec:hntt}

We utilize the navigation task from previous work~\cite{devlin2021navigation} and instantiate in the same modern AAA video game for our experiments. 
We first describe the game in more detail, then provide an overview of the navigation task. 

\subsection{The Video Game} 
To enable the reuse of agent and human-generated videos in our study, we choose the same game as previous work. 
This game is a multiplayer online combat game that features 13 customizable characters, each with special abilities.
The game is commonly compared with other popular team-based action games, such as Overwatch and DotA. 
Players compete against one another in two teams of four. 
The game has two game modes. 
One mode requires capturing and defending specific locations (called objectives) on the map, while the other involves collecting items called cells and deposit them to active platforms on the map. 
The game's team-based mechanics, objective balancing, and character customization offer a distinct multiplayer experience, making it an excellent choice for studying both AI behavior and human-AI interactions.

Underlying the game is the crucial mechanic of goal-directed navigation: players must move from one location to another to collect powerups or cells, go to drop-off platforms when they are active, and engage in combat with other players. 
As a result, navigation between points represents an abstraction of the most common task in the game. 
To allow us to concentrate on characteristics specific to navigation, we utilized a simplified version of the game that excludes other complex mechanics and objectives.

\subsection{The Navigation Task}
We instantiate the navigation task in the same way as prior work: a single avatar must navigate to a target location. 
The left screenshot of Figure~\ref{fig:be-screenshot} shows this location, indicated by the three blue containers. 
Navigating to a goal is a subtask of the main game, in which players must balance navigating to target locations to collect cells or boost health while warding off other players. 

Before the player moves, the navigation target spawns uniformly at random in one of 16 possible locations, denoted by the green crosses in the right-hand image of Figure~\ref{fig:be-screenshot}. 
Then, the player spawns on an island outside the main map (shown in the bottom portion of the mini-map) and must jump to the map's main area using the available jump areas. 
Once the player is in the central region, they can move to the target location. 

The HNTT asks human judges to identify which of two navigation behaviors more closely resemble how people navigate in \textit{reality}.
This phrasing aims to capture how \textit{convincing} an agent is~\cite{livingstone2006turing}, in contrast to another interpretation of the Turing test: whether a human or AI agent \textit{controls} an entity.
We chose this phrasing because we want to create \textit{convincing} NPCs that contribute to an immersive game experience.
In contrast, we do not wish to deceive the player into thinking that an agent is controlled by a person when it is not.

\subsection{The Baseline Agents}
Previous work~\cite{devlin2021navigation} conducted their study with two agent types: a \emph{symbolic} and a \emph{hybrid} agent.
When presented with the two agents, participants accurately detected human players above chance, meaning that people did not perceive their behavior as sufficiently human-like. 
We utilize these agents as baselines in our experiments, so we describe their essential details.

To progress toward the goal location, the agents take actions from a prespecified set (called an action space).
This action space consists of 8 possible actions: do nothing, move forward, and move left and right (30, 45, and 90 degrees on each side).
To facilitate training, the agents receive a dense reward signal to encourage successful navigation to the goal.
It consists of the following terms: 
a -0.01 per-step penalty to encourage the agent to efficiently reach the goal, 
a -1 one-time penalty for dying because the agent may fall off the map, 
an incremental reward for approaching the goal, 
and a +1 reward for reaching the goal.
We observed that this reward signal only includes terms to encourage successfully reaching the goal as quickly as possible.

The main difference between these two agents is the observations that they take as input. 
The \emph{symbolic} agent receives only a semantic, low-dimensional representation as input; the \emph{hybrid} agent also receives an image input. 
For more details about the baseline agents, we refer an interested reader to Appendix \ref{sec:app_architecture} and Devlin et al.~\cite{devlin2021navigation}.
\section{Building a More Human-Like AI}
\label{sec:hlagent}

To help design our \emph{reward-shaping} agent, we analyze the \emph{hybrid} and \emph{symbolic} agents to find characteristics that may have influenced the previous judgments of human likeness.
Based on this analysis, we introduce a novel agent for the HNTT: the \emph{\rs} agent. 

\subsection{Designing our \emph{Reward-Shaping} Agent}

This agent extends the \emph{hybrid} agent with two critical changes to promote learning of human-like behavior. 
Specifically, we introduce additional terms to the reward signal and expand the action space available to the agent.
To test whether our contributions result in differences in perceptions of human likeness, we fix all other components of our \emph{\rs{}} agent to be the same as the \emph{hybrid} agent.

Because the \emph{symbolic} and \emph{hybrid} agents previously exhibited non-human-like behavior, we inspected examples of their generated navigation and isolated three classes of problematic behavior. 
Agents would: 
\begin{description}
   \item[P1.] Wildly swing camera angles or make sudden turns,
   \item[P2.] Frequently collide with walls, and
   \item[P3.] Sometimes move more slowly than expected.
\end{description}
\pttwo{To correct these behaviors, we utilize reward shaping~\citep{rosenfeld2018leveraging} by including terms corresponding to desired or undesired behavior.}
We introduce the following terms.
First, we include a camera angle difference penalty for swift camera angle changes over a set 0.15 difference threshold value to combat \textbf{P1}.
Second, we introduce a penalty of -0.05 for any wall collisions to address \textbf{P2}.
Third, to address \textbf{P3}, we provide a penalty of -0.01 if the distance traveled between steps is lower than an environment-specific threshold value of 220 map units. 
We choose these values in line with previous training rewards and expert assessments of the relative importance of each of the components.

To encourage smoother control and avoid abrupt turns, we utilize an approach similar to action-space shaping~\cite{kanervisto2020action} by introducing additional available actions to the agent. 
Intuitively, we anticipate that the introduction of finer-grain controls will yield more fluid navigation.
We extend the action space to 14 actions from the previous 8.
In addition to the `do nothing' and `move forward' actions, we include 6 degrees of turning left and right, rather than the 3 used by the baselines.
The updated list of turning degrees for this agent is: 18, 36, 45, 54, 72, and 90 on each side.

\pttwo{Taken together, these two components comprise the novel aspects of the \emph{\rs} agent.
We design this agent in a relatively \textit{agnostic} way to make it more accessible to those without expertise in deep reinforcement learning.
Consequently, these two components can be applied to any state-of-the-art deep reinforcement learning algorithm. 
Depending on the underlying algorithm, the specific values, particularly those used for each term of the reward signal, may need to be set differently.
However, we believe that adjusting these values is more intuitive than specifying complex parameters that are specific to a particular algorithm.}

\subsection{Producing High-Quality Navigation}

\begin{figure}[t]
\centering
\includegraphics[clip,width=.99\columnwidth]{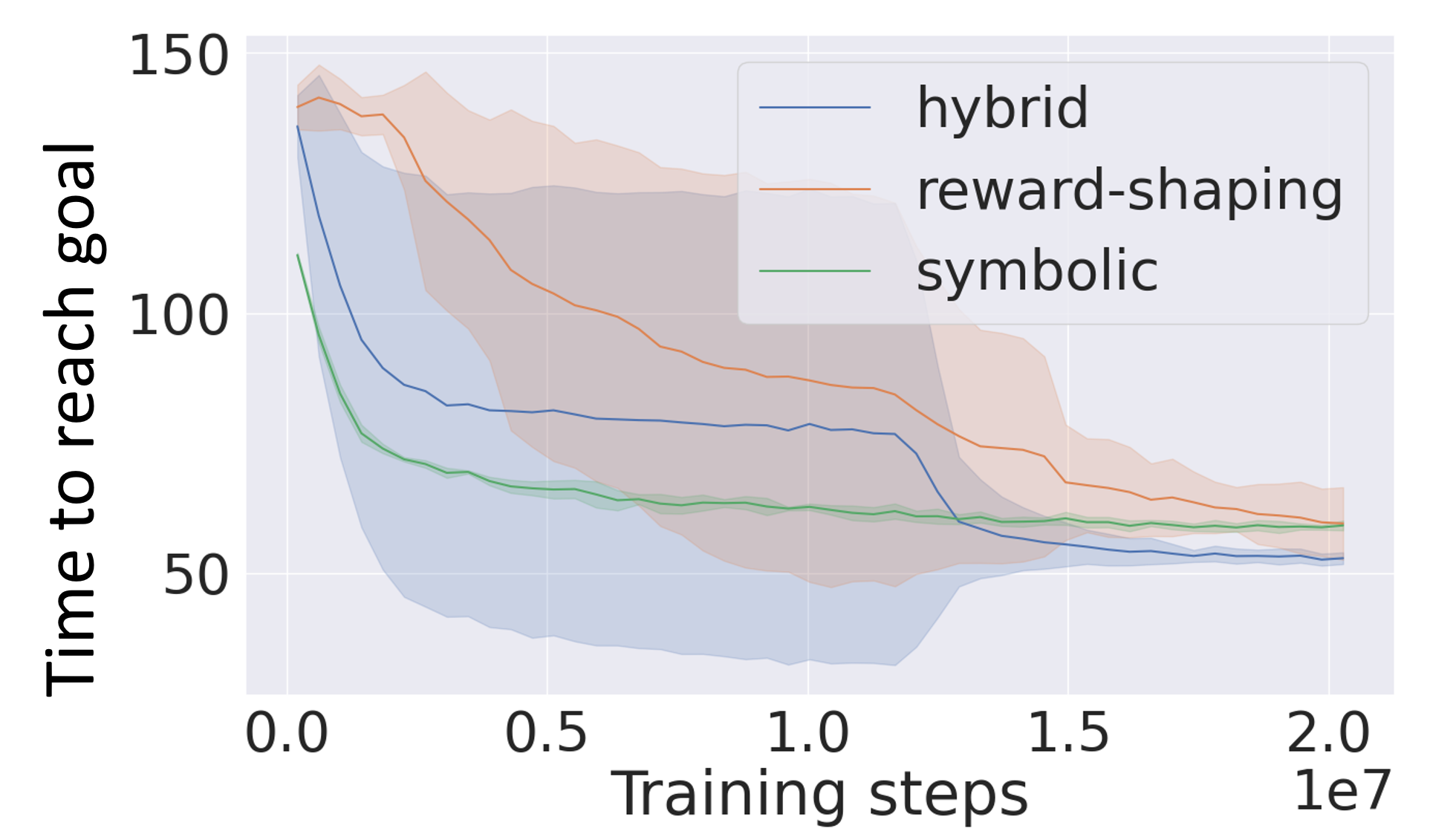}
\caption{\textbf{Hybrid, symbolic, and reward-shaping agents successfully learn to navigate.}
This plot shows the average amount of time needed to solve the task (y-axis) as a function of the amount of time taken to train the agent. 
The shaded area shows the standard deviation.
For reward-shaping, N=3; for hybrid and symbolic, N=4). 
All curves are smoothed with a rolling window of 200. 
Importantly, on average, all agents converge to solve the task in around 60 steps (around 12 seconds of in-game time).
In contrast, agents start out needing around 140 steps (around 28 seconds of in-game time) on average to solve the task.
The starting performance on this task is similar to how long an agent taking random actions would take to solve it.
The main takeaways are that performance differences are not responsible for perceived differences in human likeness, and standard metrics of task performance are insufficient to assess human likeness. 
}
\Description{Line plot showing the convergence of all three agents (hybrid, symbolic, and reward-shaping. The plot shows that on average, all agents converge to solve the task in around 60 steps per episode. Agents start out needing around 140 steps per episode on average to solve the task.}
\label{fig:convergence}
\end{figure}

We train all agents to achieve a similar level of performance on the navigation task (see Appendix \ref{app:rl_training} for the details of our training setup) to ensure that task skill is not responsible for the perceived differences in human likeness.
We measure task proficiency using the number of steps needed to reach the goal. 
Each step corresponds to around 5 seconds of real-time play.
\pttwo{Figure~\ref{fig:convergence} confirms that the agent models are indeed representative of state-of-the-art techniques for learning navigation in complex, 3D games.}

The \emph{reward-shaping} and \emph{hybrid} agents exhibit higher variance during training than the \emph{symbolic} agent. 
Because these agents must also learn from pixels, their learning task is more challenging than the \emph{symbolic} agent (that only takes in symbolic input). 
As a result, we expect higher variance during training as the agent learns this more complex task.
Importantly, all agents learn to reliably reach the goal, indicated by the performance near the end of training. 
A skilled agent now takes approximately 60 steps to complete the task (about 12 seconds of real-time play). 
This result ensures that differences in the human-likeness of assessments are not due to differences in the ability of the agents to solve the task.
\section{Experimental Design}
\label{sec:design}

To understand what characteristics people believe are indicative of human likeness, we conducted a behavioral study with human participants.
Our setup closely follows prior work~\cite{devlin2021navigation}; however, we introduce important extensions, including collecting assessments from a greater number of participants using a crowd-sourcing platform (MTurk) and additional data for a more thorough analysis.  
For completeness, we detail the full study design here.

\subsection{Experimental Task}
\label{subsec:task}

We asked each human participant to act as a judge by completing a survey consisting of \numtotalhntt{} HNTT trials.
In each HNTT trial, the judge was presented with two side-by-side video stimuli of people or agents completing the navigation task.
After watching these videos, the judge answered three questions to indicate which video they believed navigated more like a human would in the real world, a justification of their response, and an indication of their certainty.
More specifically, participants answered the following questions:
\begin{enumerate}
    \item \textbf{Which video navigates more like a human would in the real world?} The judge clicked the button underneath the video that they believed navigated more like a human would. This decision was a forced binary choice.
    \item \textbf{Why do you think this is the case? Please provide details specific to the video.} The judge answered this question as a free-form response in the box below the question. 
    \item \textbf{How certain are you of your choice?} The judge answered this question on a 5-point Likert scale, with choices ranging from extremely certain to extremely uncertain. 
\end{enumerate}

To mitigate subject learning effects from sequentially viewing multiple videos, we did not reveal to the judges which of the videos was AI-generated.
In other words, participants completed each task and, in the end, did not know which videos were human-generated. 

\subsection{Experimental Procedure} 
\begin{table}[t] 
\centering
\begin{tabular}{ccc} 
 \toprule
 Study  & Number of     & Number of \\ 
        & participants  & trials \\
 \midrule
    Human vs. \emph{hybrid} & 50 & 6  \\ 
    Human vs. \emph{symbolic} & 50 & 6 \\ 
    Human vs. \emph{reward-shaping} & 92 & 6 \\ 
 \bottomrule
\end{tabular}
\vspace{1em}
\caption{\textbf{Conditions tested in each study and the number of trials per condition.} Importantly, note that the human vs. \emph{hybrid} and human vs. \emph{symbolic} studies are replications of prior work \cite{devlin2021navigation} to validate the switch to a crowd-sourcing platform. 
}
\Description{Table showing the different conditions tested in each study. The table provides an overview of the number of participants in each study, the number of trials, and the conditions tested.}
\label{table:hntt-conditions}
\end{table}

\begin{figure*}[t]
\resizebox{.9\textwidth}{!}{
\centering
\fbox{
  \subfloat[]{\label{fig:pre-questions}\includegraphics[width=.5\textwidth]{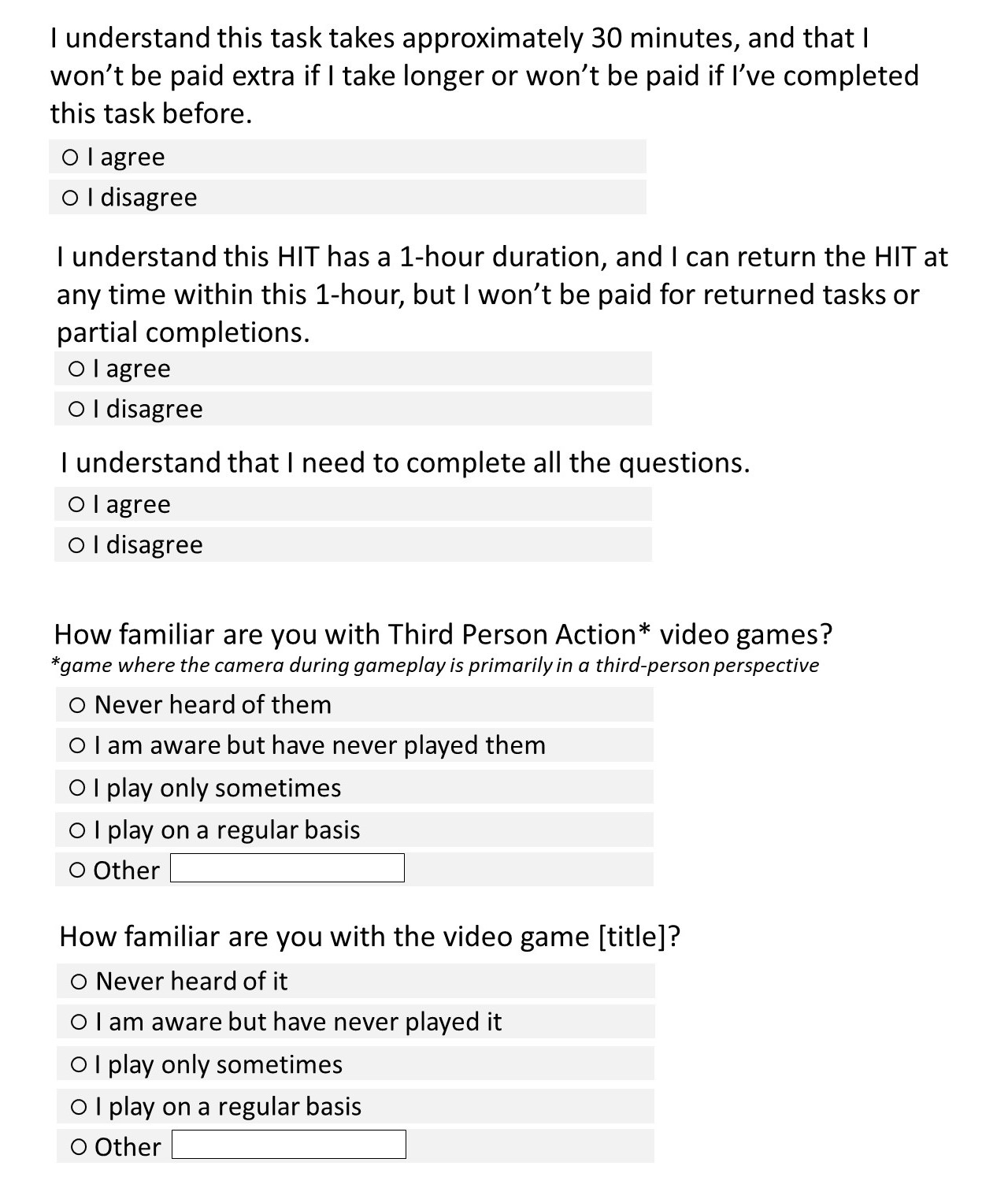}} \rulesep
  \subfloat[]{\label{fig:hntt-trial}\includegraphics[width=.5\textwidth]{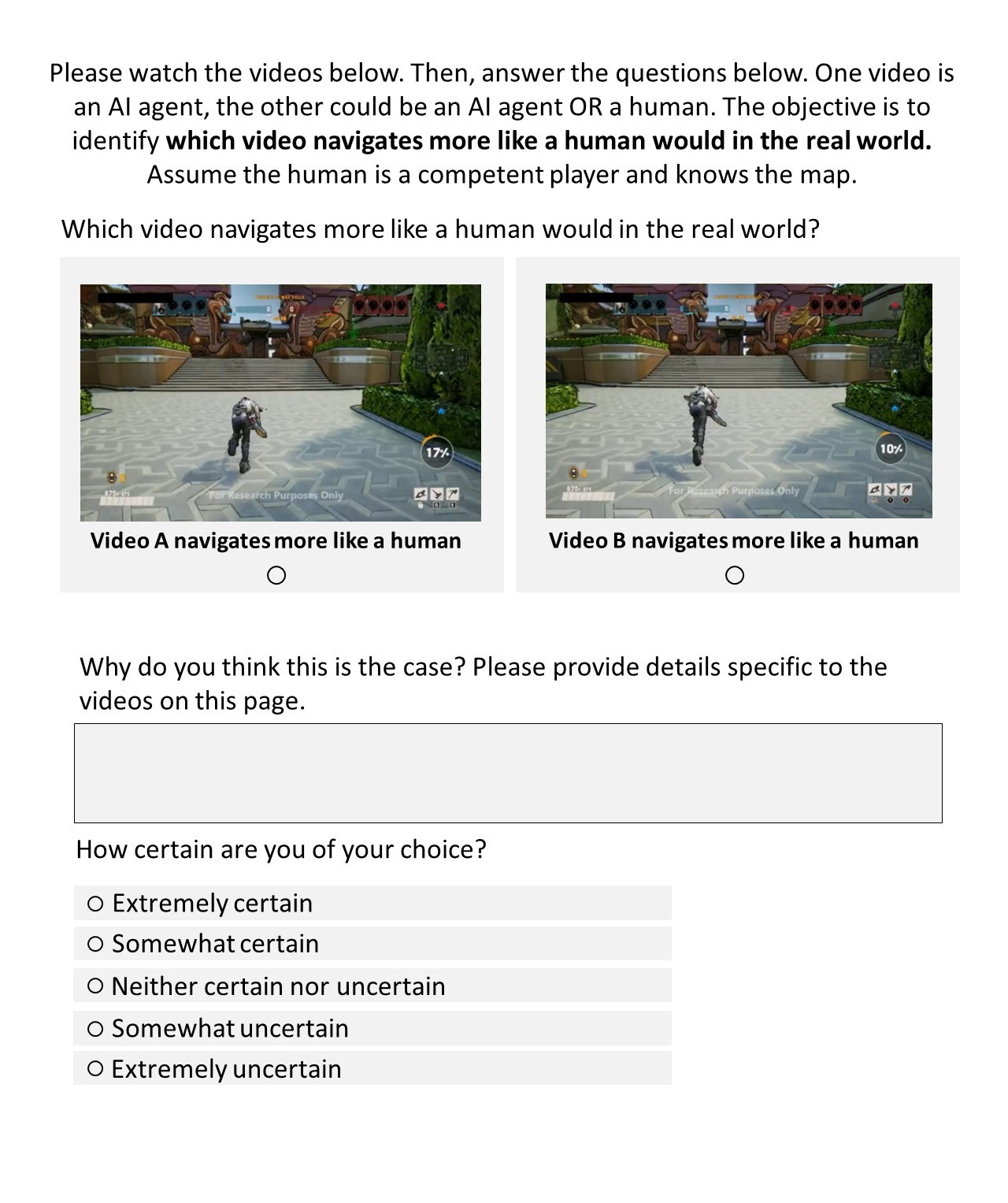}}
}
}
\caption{\textbf{Screenshots of HNTT survey questions.} The screenshot in (a) shows the comprehension and familiarity questions (asked once per participant). 
We gauge the participant's familiarity with the time the task will take, understanding of task completion, familiarity with third-person action video games, and familiarity with the video game used in the survey. The screenshot in (b) depicts one HNTT trial.
We ask participants to choose their response to the human likeness question, justify it, and indicate their level of certainty.}
\Description{Screenshot of HNTT survey questions to gauge the participant's familiarity with the time the task will take, understanding that they need to complete all questions, familiarity with third-person action video games, and with the video game used in the survey. Another screenshot shows a single HNTT trial.}
\label{fig:hntt_survey}
\end{figure*}

We completed 3 studies; each study pitted a human-controlled agent against a different AI agent.
Within each study, all judges viewed the same \numtotalhntt{} trials.
The trials were presented in a randomized order per judge. 
Within each trial, the ordering of the two videos was randomized, such that the human-generated video could not be inferred by presentation order.
Table \ref{table:hntt-conditions} outlines the conditions tested in each study.

Each participant first read through an introduction page with the required task instructions (see Appendix \ref{sec:app_study} for the full text). 
They then completed a consent form and read through a background page with brief details about the video game.
They answered a series of questions to assess their comprehension of the task and familiarity with video games. 
Finally, participants engaged in the \numtotalhntt{} HNTT trials.
Figure \ref{fig:hntt_survey} shows screenshots of the comprehension and familiarity questions (a) and an example HNTT trial (b).


\subsection{Navigation Video Generation and Sampling}
A key part of the study is the videos that were shown to the human judges.
For the human-generated navigation data and videos, we use the publicly-available sample published by previous work~\citep{devlin2021navigation}.\footnote{\ptthree{Data use under MSR-LA license. License details can be found in the original authors’ GitHub \url{https://github.com/microsoft/NTT}. This link includes \textit{all} data used in this study, including from our \textit{reward-shaping} agent.}}
We sampled human videos from the 40 published under their ``study 1'' protocol. 
To generate the AI navigation data, we select each agent's most recently saved version. 
Then, we instantiate a new session and deploy the agent in the game 100 times, producing 100 total navigation videos per agent.
To produce the video stimuli used in the study, we sample the recordings uniformly at random.

\ptthree{
We implemented several measures to standardize the videos and ensure that any measurement noise applied to all conditions.
First, we checked that any changes in light applied similarly across conditions. 
Second, we designed the timing of the stimuli to ensure that participants had sufficient time to engage in and provide meaningful responses in all trials.
As a result, we did not use videos that were too long and excluded videos shorter than 10 seconds (before post-processing) because they were deemed too short to assess navigation quality in pilot studies. 
Third, because the goal locations may differ depending on the game-controlled initialization, we matched the goal locations of the human videos with the AI agent videos.
Consequently, we used different human videos for different studies.
Fourth, we applied the post-processing steps from prior work \citep{devlin2021navigation}), including masking identifying information, adding a ``For Research Purposes Only" watermark, and cutting out the last few seconds of the human videos.
We implemented the last change to correct an effect of the data collection process, where the human players manually ended their recording, adding a few seconds at the end of the videos.}

\subsection{Other Experimental Control} 
The MTurk crowd-sourcing platform \cite{paolacci2010running} is widely used for data collection and research due to its scalability, as long as researchers implement appropriate steps for quality control~\cite{ipeirotis2010quality}.
Here, we detail the study inclusion criteria that we implemented for quality control.

We set the following MTurk requirements for survey participation: location is United States, age is 18 or older, and language is English. 
We did not collect demographic information or any other personally identifiable information. 
To target more experienced MTurk Workers, we set the following Human Intelligence Task (HIT) qualifications: HIT Approval Rate greater than 98\%, Number of HITs Approved greater than 500, and a qualification to prevent repeat responses. 
To incentivize quality, we included a bonus payment for each high-quality response. 
We reviewed the free-form answers to find low-quality or suspected bot responses; for example, we excluded from analysis responses with high instances of typos, copy/pasted answers, or nonsensical wording. 
We paid all participants who completed the task for the HIT, even if their response was identified as low-quality. 
The low-quality responses did not receive the bonus payment. 
We paid on average 15 USD per hour. 
We obtained approval for our studies from our Institutional Review Board (IRB) and informed consent from each participant. 
We included details of the study and a description of any potential participant risks in the consent form. 

\begin{table*}[t]
\centering
\begin{tabular}{
|c|l|l|l| 
}
  \hline
  Response & Free-Form Response & More Human-Like & Less Human-Like  \\ 
  \hline
   B & The character in Video B runs in \hlblue{straight lines} and \hlblue{goes to where} & Smoothness of movement $+$; & Collision avoidance $-$; \\
  & \hlblue{he needs to be going}. The character in Video A is \hlblue{running in} & Goal directed $+$ & Goal directed $-$ \\
  & \hlblue{circles},\hlblue{into objects}, etc. & & \\
  \hline
\end{tabular}
\caption{\textbf{Example coded response to the question, "Which video navigates more like a human would in the real world?"}. 
The leftmost column indicates that this judge believed Video B to exhibit the more human-like behavior. 
The highlighted text illustrates the annotation process. 
The judge identifies that the more human-like character runs in straight lines (more human-like code: smoothness of movement $+$) and navigates to the goal (more human-like code: goal directed $+$), while the character that they believe is less human-like runs in circles (less human-like code: goal directed $-$) and into objects (less human-like code: collision avoidance $-$).}
\label{table:coded-response}
\Description{Table showing an example coded response to HNTT trial question. The judge identifies that the more human-like character runs in straight lines (more human-like code: smoothness of movement $+$) and navigates to the goal (more human-like code: goal directed $+$), while the character that they believe is less human-like runs in circles (less human-like code: goal directed $-$) and into objects (less human-like code: collision avoidance $-$).}
\end{table*}

\section{Analysis}
Our primary objective is to evaluate the human-likeness of the agents using both quantitative and qualitative measures. 
To quantify the ability of the human judges to distinguish between the human-like and non-human-like agents, we analyze their accuracy scores and self-reported uncertainty. 
To identify the factors that influence their perceptions of human likeness, we adopt a qualitative approach. 
We construct and use codes to summarize the reasons cited in the open-ended responses and compare the frequency of these codes across different settings.

\subsection{Assessing Human-Likeness}
\label{subsec:hntt-criterion}
We first aim to identify which agents pass the HNTT according to our proposed criterion.
Because existing work demonstrates differences in assessment ability depending on expertise, we seek to identify whether this phenomenon holds in our setting.
We finally seek to investigate the relationship between self-reported uncertainty and accuracy when assessing the agents.
We instantiate the following research questions:
\begin{description}
   \item[RQ 1.] Which agents are judged as being human-like?
   \item[RQ 2.] Do the judges exhibit greater accuracy in assessing human likeness as a function of their experience with games?
   \item[RQ 3.] What is the relationship between the accuracy of human judges and their self-reported uncertainty?
\end{description}

To answer \textbf{RQ 1}, we propose a firm criterion for deciding whether an agent is sufficiently human-like, formalizing the question: \emph{are human assessors unable to distinguish between agent and human behavior}?
We implement this criterion as a statistical test that determines whether human judges distinguish between human and agent behavior at a level significantly different from chance. 
We instantiate this test by computing the 95\% confidence interval for the median of the human-agent comparisons using bootstrap sampling (a non-parametric approach). 
If the 95\% confidence interval includes 0.5 (chance-level agreement), then the agent passes the HNTT.

For both \textbf{RQ 2} and \textbf{RQ 3}, we compare our variables of interest with \textit{accuracy}.
We define accuracy to mean that the participant identified that the human-generated behavior was more human-like than the AI-generated behavior.
To answer \textbf{RQ 2}, we compare accuracy to the self-reported familiarity of the participants with action games in general and the specific game in the study.
To answer \textbf{RQ 3}, we examine the self-reported uncertainty of the judges and its relationship to accuracy.

\subsection{Assessing Human-Like Characteristics}
\label{subsec:data_analysis}
\begin{table*}[t]
 \centering
 \begin{tabular}{ccccc} 
    \toprule
   \textbf{Annotation Code} & \textbf{Shorthand} & \textbf{Definition} & \textbf{Key Words and Phrases} & \textbf{Example Snippet} \\ 
   \hline
   Smoothness of & smooth & The quality of the agent's navigation & Smooth, jerky, straight, & Movements are way \\
   movement &  & or camera movement & swerve, steady, fluid & more smooth \\
   \hline
   Goal directed & goal & How goal-directed the agent's behavior & Intention, focus, & Deliberate camera \\
   & & seems & knew where to go & movements \\

   \hline
   Collision & avoidance & Whether the agent avoids collisions & Collide, avoid, crash & Runs into a box \\
   avoidance & & & runs into obstacle &  \\  
   \hline
   Environment & receptivity & Whether the agent understands and/or & Explore, stay on path, & Ignores all the  \\ 
   receptivity  & &  properly interacts with the environment  & collect power-ups & health/mana/etc \\ 
   \hline 
   Intuition & intuition & The judge cannot pinpoint behaviors & Natural, feeling, & Just a feeling \\
   & & & seems to be & \\
   \hline 
   Self-reference & self-reference & Relationship to the judge's own & Like I play & [Like] how I navigate  \\ 
   & & movement or play &  & with that ... view \\ 
   \hline
 \end{tabular}
 \caption{\textbf{Annotation code definitions.} 
 The codes used to label the free-form responses are presented in the leftmost column. 
 The middle-left column shows the corresponding shorthand for the codes, used later in the paper.
 In the middle column, a brief definition of each code is presented. 
 The middle-right column lists the keywords and phrases that the annotators used to determine if a response could be labeled as containing a particular code. 
 An example snippet of a response that would be labeled with that code is provided in the rightmost column. 
 Although the included examples are fairly clear, the free-form responses often contain more ambiguous content.
 }
 \label{table:test-coded-response}
     \Description{Table describing all annotation codes used, as well as their definitions.}
 \end{table*}

 To analyze the \textit{characteristics} that correspond to assessments of human likeness, we instantiate the following research questions:
\begin{description}
   \item[RQ 4.] Are there key differences between how people characterize human-like and non-human-like behavior? Does this differ when the agent does or does not pass the HNTT?
   \item[RQ 5.] What is the relationship between the characteristics that people use to assess human likeness and their ability to accurately assess it?
\end{description}

We selected a sub-sample of the responses from the \studyone{} and the \studythree{} studies for analysis.
We chose these studies to enable comparison between an agent that does not pass the HNTT with one that does (see Section~\ref{subsec:results-hntt}).  
We first randomly sampled a set of $55$ responses to compute the initial agreement, called the \textit{agreement sample}.
We filtered this sample to $53$ after removing responses that were ambiguous or could not be categorized by any of our codes.
We then constructed the sample for analysis by randomly sub-sampling three free-form responses per judge for each study.
To minimize bias, we shuffled responses before sampling.
We removed responses that were ambiguous or could not be categorized by any of our codes, resulting in a dataset of $395$ responses for our analyis of human-like characteristics.

We followed a pair-coding approach to annotate the data. 
The annotator with more familiarity with the data proposed an initial list of codes derived from previous work~\cite{zuniga2022humans}.
Following established notation~\cite{artstein2008inter}, we have a set of $I$ items (or responses), labeled as at least one of the $K$ categories by $C=2$ coders.
We decompose each label as more or less human-like $H=(\texttt{more, less})$ and quantify its direction $D=(\texttt{more, less})$, when applicable. 
For example, if we label item $i$ as \textit{smoothness of movement}, we note whether the judge considered the behavior human-like and whether they noted it as being more $+$ or less $-$ smooth.

The two annotators then convened to discuss the meaning of the codes and jointly code a set of $5$ responses. 
Table \ref{table:coded-response} illustrates an example of a coded response.
After that, the two annotators separately coded the agreement sample with the initial set of codes. 
Optionally, the annotators could label responses as \textit{other} and provide specific examples to enable revisions of the codes if other themes emerged.
The two annotators iteratively reconvened to discuss disagreements and refine the codes.
After multiple rounds of discussion, independent coding, and disagreement resolution, the annotators fixed the set of codes (Table \ref{table:test-coded-response}) and their inclusion criteria to label the full sample.

Because we aim to design human-like AI agents, we want to identify codes that could be utilized by AI designers. 
For that reason, when deciding on codes, we prioritize codes that refer to specific behaviors over more general ones.
For example, a collision avoidance behavior could be coded as goal-directed; however, we code it only as collision avoidance.
This protocol promotes the independence of categories while prioritizing specific, lower-level behaviors to use in designing agents.
When coding, the annotators first consider whether the response could be categorized as a lower-level code, then move to more general codes if needed.
Appendix~\ref{app:coding} contains more details about this process.

 \begin{table}[t]
 \centering
 \begin{tabular}{cccc} 
 \toprule
   \textbf{Annotation} & \textbf{Direction} & \textbf{Cohen's $\kappa$} & \textbf{Cohen's $\kappa$ } \\ 
   \textbf{Codes} & & \textbf{Agreement} & \textbf{Overlapping} \\  
    & & \textbf{Sample} & \textbf{Sample} \\
   \hline
   Smoothness of & More + &  0.90 & 0.79 \\
    movement & Less - &  0.64 & 0.64 \\
   \hline
   Goal directed &More + & 0.82 & 0.78 \\
   & Less - & 0.82 & 0.63 \\
   \hline
   Collision  & More + & 1.00  & 1.00\\
   avoidance & Less - & 0.64  & 0.73\\
   \hline
   Environment & More + & 0.82 & 0.93 \\
    receptivity & Less - & 0.73 & 0.67 \\ 
   \hline 
   Intuition & & 1.00 & 0.87 \\ 
   \hline 
   Self-reference && 1.00 & 1.00 \\
   \midrule
   & Average & \textbf{0.84} & \textbf{0.78} \\
   \hline 
 \end{tabular}
 \caption{\textbf{Per-code Cohen's $\kappa$ score.} The two annotators achieved an average Cohen's $\kappa$ score of $0.84$ over all of the codes for the \textit{agreement sample}. 
 According to Cohen's suggested interpretation, we achieve at least moderate agreement on each category and achieve almost-perfect agreement on $7$ of the $10$ categories when annotating the agreement sample. 
 When annotating the \textit{overlapping sample}, the two annotators achieved an average Cohen's $\kappa$ score of $0.78$ over all of the codes. 
 According to Cohen's suggested interpretation, we achieve at least substantial agreement on each category.
 There was only a small overall decrease in agreement between these two settings, indicating that our coding process is fairly general.
 }
 \Description{Table showing Cohen's kappa calculations for the agreement and overlapping samples. The table shows that, on average, the annotators achieve high agreement in both cases.}
\label{tab:kappa}
 \end{table}

The annotators achieved an overall average inter-annotator agreement of $0.84$ on the agreement sample.
We calculate inter-annotator agreement with binary Cohen's kappa $\kappa$ \citep{cohen1960kappa} over $K$, $D$, and $H$, as previously defined. 
See Table \ref{tab:kappa} for more details.
After fixing the list of codes, the annotators divided the data sub-sample such that there was overlap on 25\% of the data (99 items).
We report Cohen's kappa in Table \ref{tab:kappa} for the overlapping sample to ensure that our understanding of the codes did not overfit the specific examples in the agreement sample.

We provide a more detailed discussion of the annotation codes and inclusion criteria. 
Table \ref{table:test-coded-response} includes these definitions and phrases that helped us identify the presence of each code.
For each code, we provide a supporting example to give the reader a sense of what common responses may look like.
\emph{Smoothness of movement} refers to the quality of the agent's navigation or camera movement. 
This code considers both immediate jerky actions and temporally-extended zig-zagging behavior. 
\emph{Goal directed} refers to how intentional the agent's behavior appears. 
We include descriptions of behavior that pertain to a perceived goal, even if that goal is not the primary one. 
We include the code \emph{collision avoidance} because it is a long-standing area of research in the robotics community~\cite{seiler1998development}. 
This code refers to intentional behavior to redirect from a potential crash. 
\emph{Environment receptivity} aims to capture the agent's relationship with the game environment, its contextual understanding, and adherence to norms. 
In a real-world setting, this might look like a person walking on a path instead of the grass or crossing the street when permitted by a pedestrian signal.  
Any responses that refer to non-specific feeling that a behavior was more human-like are categorized as \emph{intuition}. 
We include this code to capture instances where participants can identify what they believe is more human-like behavior but struggle to express it.
Finally, we include \emph{self-reference} as a code to capture when judges relate the agent's behavior to their own play.

During the iterative coding process, the annotators assessed the likely causes of disagreements.
After resolving mistakes and other easy-to-resolve issues, the annotators determined that the remaining disagreements arose from individual differences in interpreting ambiguous natural-language responses. 
This cause means that neither annotator can be treated as more correct for disagreement resolution.
The annotators, therefore, decided on the following disagreement resolution scheme. 
When a disagreement arises in at least one label for an item annotated by both annotators, we randomly choose an annotator to treat as correct and use their labels. 
\section{Results}
\label{sec:results}
We first present the results from our analysis described in Section \ref{subsec:hntt-criterion}; in particular, we demonstrate that our reward-shaping agent passes the HNTT while other agents do not.
We then present the results from our analysis described in Section \ref{subsec:hntt-criterion} by highlighting characteristic behaviors and key differences in how human judges perceive AI vs human players. 
We find that people tend to utilize similar high-level characteristics when characterizing human-like behavior. 
However, their beliefs about AI capabilities may inform whether they think AI agents more or less strongly exhibit these characteristics.

\subsection{Analysis of Human Likeness}
\label{subsec:results-hntt}

\begin{table}[t]
\centering
\begin{tabular}{c l}
\toprule
Agent & Median Accuracy (IQR) [95\% CI] \\
\hline
\emph{symbolic} & $0.83$ $(0.67-1.00)$ $[0.67, 1.00]$ \\ 
\emph{hybrid} & $0.83$ $(0.67-1.00)$ $[0.83, 1.00]$ \\ 
\emph{\rs} & $0.50$ $(0.33-0.67)$ $[0.50, 0.50]$ \\ 
\hline 
Agent & Median Uncertainty (IQR) \\ 
\hline 
\emph{symbolic} & $2.17$ $(1.67-2.42)$ \\ 
\emph{hybrid} &  $1.92$ $(1.33-2.25)$ \\ 
\emph{\rs} & $2.17$ $(1.75-2.67)$ \\ 
\bottomrule
\end{tabular}
\vspace{.5em}
\caption{\textbf{Full summary statistics of accuracy and uncertainty.} 
We show the median accuracy (IQR=Q1-Q3) for each agent, reported as non-parametric measures of central tendency and spread; we report 95\% confidence interval and median uncertainty (IQR=Q1-Q3) of the human-agent comparisons for each agent. 
Only the \emph{reward-shaping} agent passes the HNTT according to our proposed metric.}
\Description{Full summary statistics showing that the reward-shaping agent passes the HNTT, while the baselines do not.}
\label{table:hntt-res2}
\end{table}

\textbf{Only the \emph{reward-shaping} agent passes the HNTT.}
Table~\ref{table:hntt-res2} shows the full summary statistics, which are computed over the full dataset from our survey.
Each bootstrap calculation is run over 10000 iterations.
The \emph{symbolic} and \emph{hybrid} baseline agents do not pass the HNTT according to our criterion. 
The judges had median accuracies of $0.83$ (\emph{symbolic} agent, 95\% CI=[0.67, 1.0]) and $0.83$ (\emph{hybrid} agent, 95\% CI=[0.83, 1.0]), indicating that they distinguish the agents from humans significantly higher than chance level.
In contrast, our \emph{reward shaping} agent passes this test of human-likeness: the median accuracy has a 95\% confidence interval that includes 0.5 (chance-level agreement).
This result suggests that the judges cannot consistently differentiate between the \emph{reward shaping} agent and the human player (\emph{reward shaping} agent, median accuracy=$0.50$, 95\% CI=[0.50, 0.50]).

Because the sample sizes of the trials differ (50 samples for the human vs. \emph{hybrid} and human vs. \emph{symbolic} conditions; 92 samples for the human vs. \emph{reward-shaping} condition), we validate our results by subsampling the data for the \emph{reward-shaping} agent to 50 samples, then run the bootstrap sampling procedure $100$ times.
We find that the computed CI always contains 0.5, or chance-level agreement, in each run of the bootstrap. 
The average median accuracy is $0.50$, with a variance of $0.00$; the averaged CI is [0.44, 0.63], with a variance of 0.01 for the lower bound and 0.00 for the upper bound. 
We, therefore, answer our \textbf{RQ 1}: the \textit{reward-shaping} agent is the only agent that is judged as human-like according to this proposed metric.

\textbf{There is no relationship between game familiarity and ability to accurately assess the human likeness of the AI agents.}
For each study, we perform a multiple linear regression analysis to test whether specific game familiarity and general game familiarity significantly predicted accuracy in assessing human-likeness. 
There is no relationship between either of the self-reported familiarities and accuracy for all agents. 
For the \emph{symbolic} agent, the fitted regression model was:
\begin{align*}
    \texttt{accuracy} & = 0.68 - 0.01 (\texttt{specific game familiarity}) \\
    & + 0.03 (\texttt{general game familiarity}).
\end{align*}
The overall regression was not statistically significant ($R^2=0.01$, $F(2,47) = 0.21$, $p=0.814$).
Decomposing the results further, neither specific game familiarity ($\beta=-0.01, p=0.878$) nor general game familiarity ($\beta=0.03, p=0.525$) predicted accuracy. 

For the \emph{hybrid} agent, the fitted regression model was:
\begin{align*}
\texttt{accuracy} &= 0.67 - 0.03 (\texttt{specific game familiarity}) \\ 
&+ 0.06 (\texttt{general game familiarity}).
\end{align*}
The overall regression was not statistically significant ($R^2=0.06$, $F(2,47) = 0.26$, $p=0.261$).
We found that specific game familiarity did not significantly predict accuracy ($\beta=-0.03, p=0.377$). 
General game familiarity also did not significantly predict accuracy ($\beta=0.06, p=0.109$).

Turning our attention to the \emph{reward-shaping} agent, the fitted regression model was:
\begin{align*}
\texttt{accuracy} &= 0.40 - 0.02 (\texttt{specific game familiarity}) \\
& + 0.04 (\texttt{general game familiarity}).
\end{align*}
The overall regression was again not statistically significant ($R^2=0.02, F(2,89) = 0.94, p=0.393$). 
This result holds for both specific game familiarity ($\beta=-0.02, p=0.522$) and general game familiarity ($\beta=0.04, p=0.191$).

These findings suggest that game familiarity is generally \textit{not} predictive of accuracy for this specific task, answering \textbf{RQ 2}.
In contrast, previous findings have demonstrated a relationship between the ability to assess human likeness and familiarity with the domain of study.
We suspect that this result differs because we are studying a relatively simple setting, in which most people have strong priors about what constitutes human likeness. 
Navigating by walking or running is an activity that most people either perform or observe daily, meaning we will likely have a strong internal sense of human-like \smnote{movement} --- even if we are not familiar with games that require navigation.
In contrast, we hypothesize that game familiarity would be predictive of accuracy in the full game setting, implicating the importance of assessing human likeness in more complex settings as an important next step.

\textbf{Human judges exhibit less false confidence in their assessments of the \emph{reward-shaping} agent.}
We assess the median uncertainties of participants; lower values correspond to more certainty and higher values correspond to less certainty.
Table~\ref{table:hntt-res2} depicts the results of this analysis.
Participants reported similar levels of uncertainty when assessing our \emph{reward-shaping} agent (median=2.17, IQR=(1.75-2.67)) and the symbolic agent (median=2.17, IQR=(1.67-2.42)).
In comparison, participants reported higher certainty when assessing the \emph{hybrid} agent (median=1.92, IQR=(1.33-2.25)).

People felt less confident about their assessments of the \textit{symbolic} and \textit{reward-shaping} agents compared to the \textit{hybrid} agent; however, participants more accurately detected human-generated behavior in the presence of the \textit{symbolic} and \textit{hybrid} agents. 
This result is surprising because it suggests that self-reported uncertainty and accurate assessments are not necessarily correlated. 
In other words, participants may exhibit false confidence in their ability to assess the human likeness of agents. 
We believe that participants may have been less certain about their assessments of the \textit{symbolic} agent due to differences in the lengths of the videos: on average, the videos of the symbolic agents were 8.3 seconds long, whereas the hybrid agent videos were 15.3 seconds long.
Participants may have not had enough time with the agent to accurately assess it.
Taken together, the accuracy and uncertainty results indicate that, when presented with behavior from the \textit{reward-shaping} agent, participants exhibited less false confidence in their assessment ability compared to when they were presented with behavior generated by the \textit{hybrid} agent.
This result answers \textbf{RQ 3}.


\subsection{Analysis of Human-Like Characteristics}
\label{subsec:results-text}

\begin{figure}[t]
\centering
\includegraphics[width=.99\columnwidth]{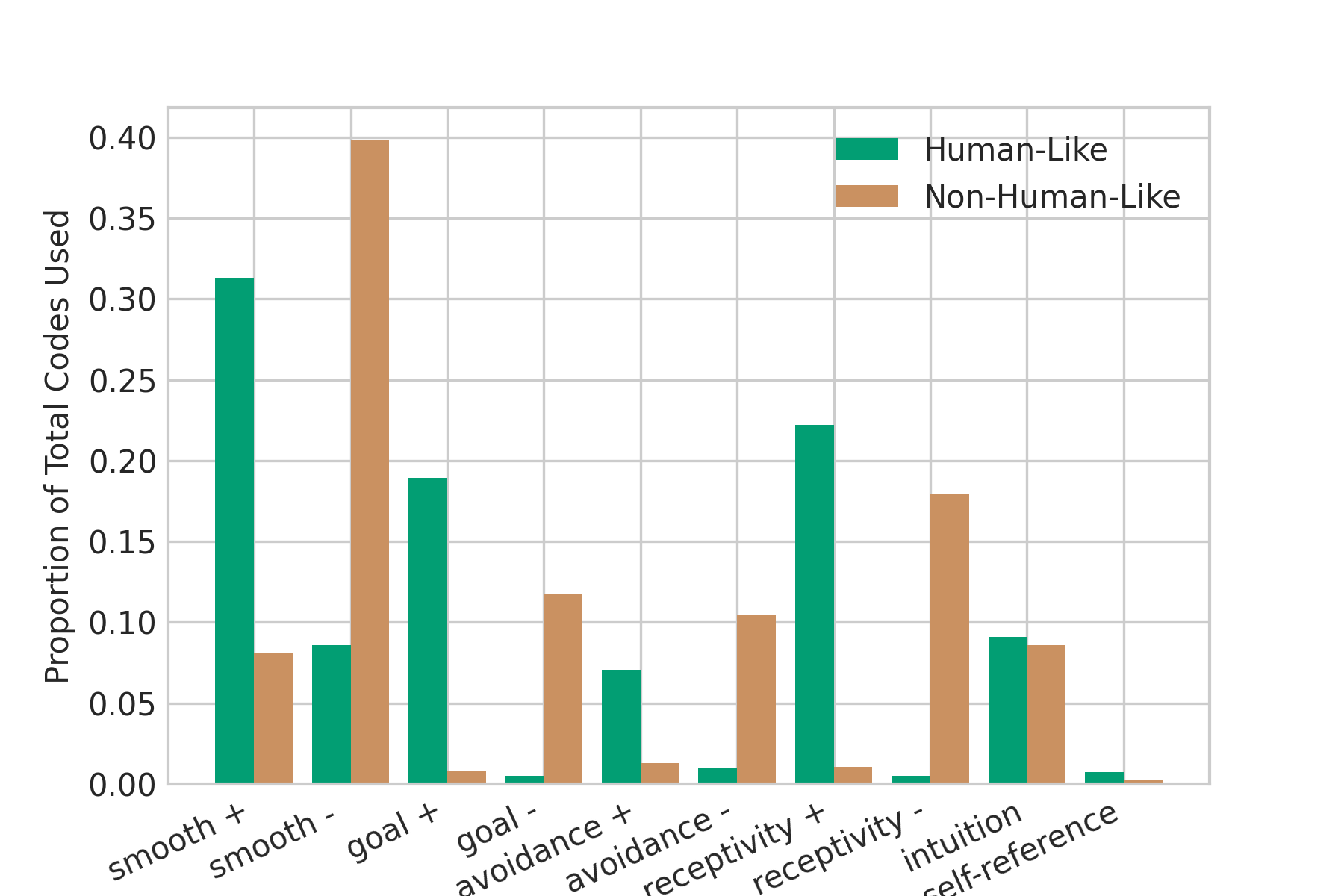}
\caption{\textbf{Codes used to describe human-like and non-human-like behavior.} We compare the proportion of codes used to describe human-like and non-human-like behavior by human judges in their assessment of human likeness. People more frequently characterize human-like behavior as being more smooth, receptive and responsive to the environment, and goal-directed. In contrast, participants more frequently describe non-human-like behavior as being less smooth, receptive and responsive to the environment, and goal-directed.}  
\label{fig:hntt-counts}
\Description{Bar plot showing the codes used to describe human-like and non-human-like behavior. The judges use similar high-level characteristics (smooth, receptive and responsive to the environment, goal-directed). However, judges more often describe human-like behavior as having more of those characteristics and non-human-like behavior as exhibiting less.}
\end{figure} 


In all plots, we use the shorthand version of the codes, noted in Table~\ref{table:test-coded-response}, along with the $+$ and $-$ notation.
The $+$ and $-$ notation indicate the degree, or direction, of the code. 
For example, smooth $+$ indicates that the participant referenced more smooth movement, and smooth $-$ indicates that the participant referenced less smooth movement. 

\textbf{Human judges rely on similar high-level characteristics when assessing human-like behavior.}
Figure \ref{fig:hntt-counts} shows the codes that participants use to describe human-like and non-human-like behavior.
We investigate the relative number of times a code was used compared to all codes used to describe either human-like or non-human-like behavior (human-like and non-human-like code proportions should sum to $1$). 
Judges tend to rely on similar high-level characteristics when characterizing human-like behavior. 
Overall, they most often reference the following high-level codes: smoothness of movement, environment receptivity, and goal-directedness. 
When we decompose the responses based on whether the behavior was assessed as human-like or not, we find that people more frequently characterize human-like behavior as more smooth, receptive and responsive to the environment, and goal-directed. 
In contrast, participants more frequently describe non-human-like behavior as being less smooth, receptive and responsive to the environment, and goal-directed.
They rely on intuition and self-reference to a similar degree when describing human-like and non-human-like behavior. 

We investigated these responses based on agent type but did not find a difference between the resulting proportions of codes. 
This result supports the assertion that people may have relatively stable beliefs what constitutes human-like behavior.
Therefore, the rationale is only sometimes useful: in other words, looking for the jerkier agent only makes sense if the AI has not been designed to be less jerky than the person. 
We, therefore, conclude that, although people rely on different specific characteristics to determine human likeness, the general characteristics are relatively stable across different AI agents, which answers \textbf{RQ 4}.

\begin{figure*}
    \centering
    \begin{subfigure}[b]{0.49\textwidth}
        \includegraphics[width=1.0\textwidth]{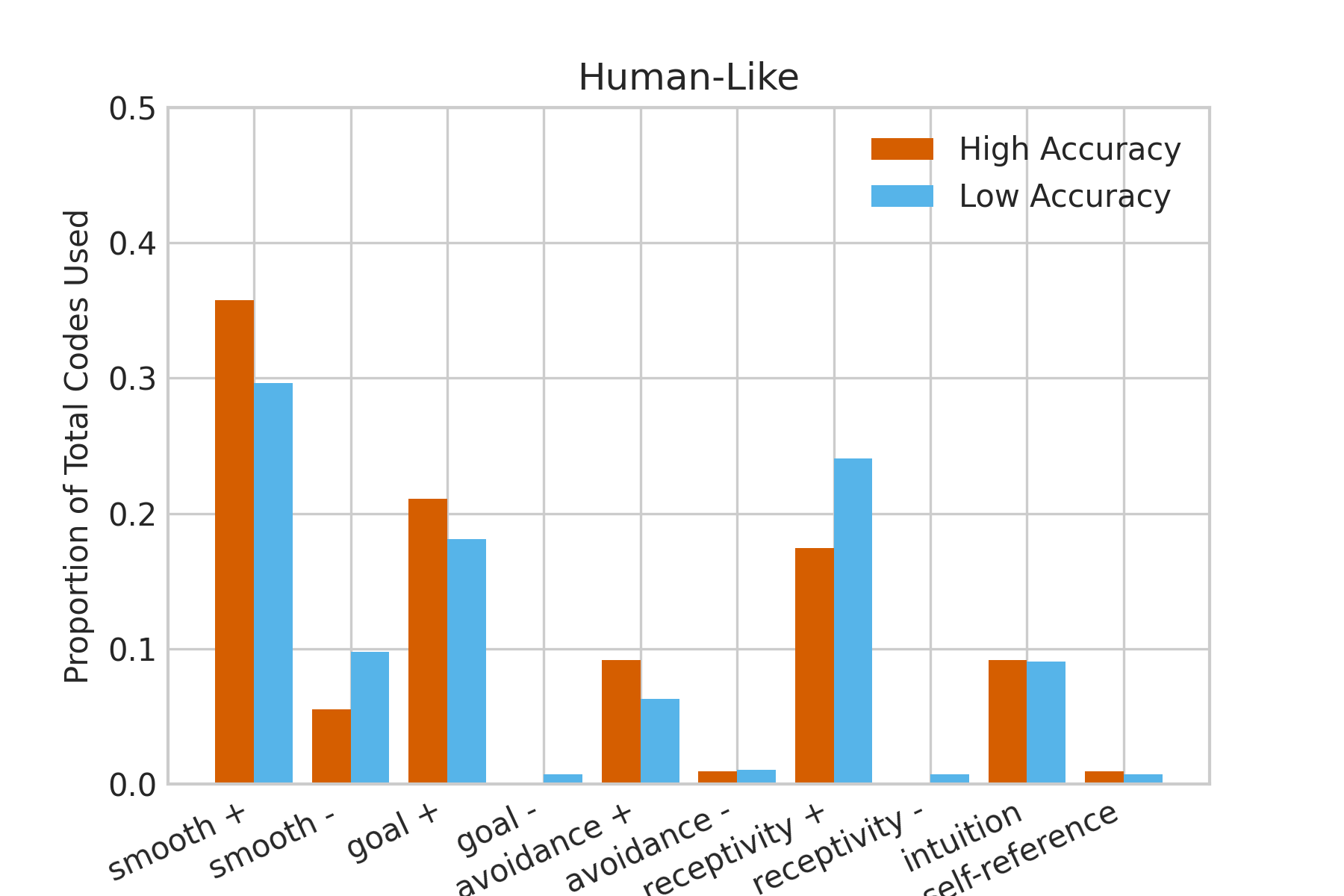}
        \caption{\textbf{Codes used to describe human-like behavior by low- and high-accuracy judges.} We compare the proportion of codes used to describe human-like behavior by human judges in their assessment of human likeness.
    }
        \label{fig:hntt-hl-acc}
        \Description{}
    \end{subfigure}
    \hfill
    \begin{subfigure}[b]{0.49\textwidth}
        \includegraphics[width=1.0\textwidth]{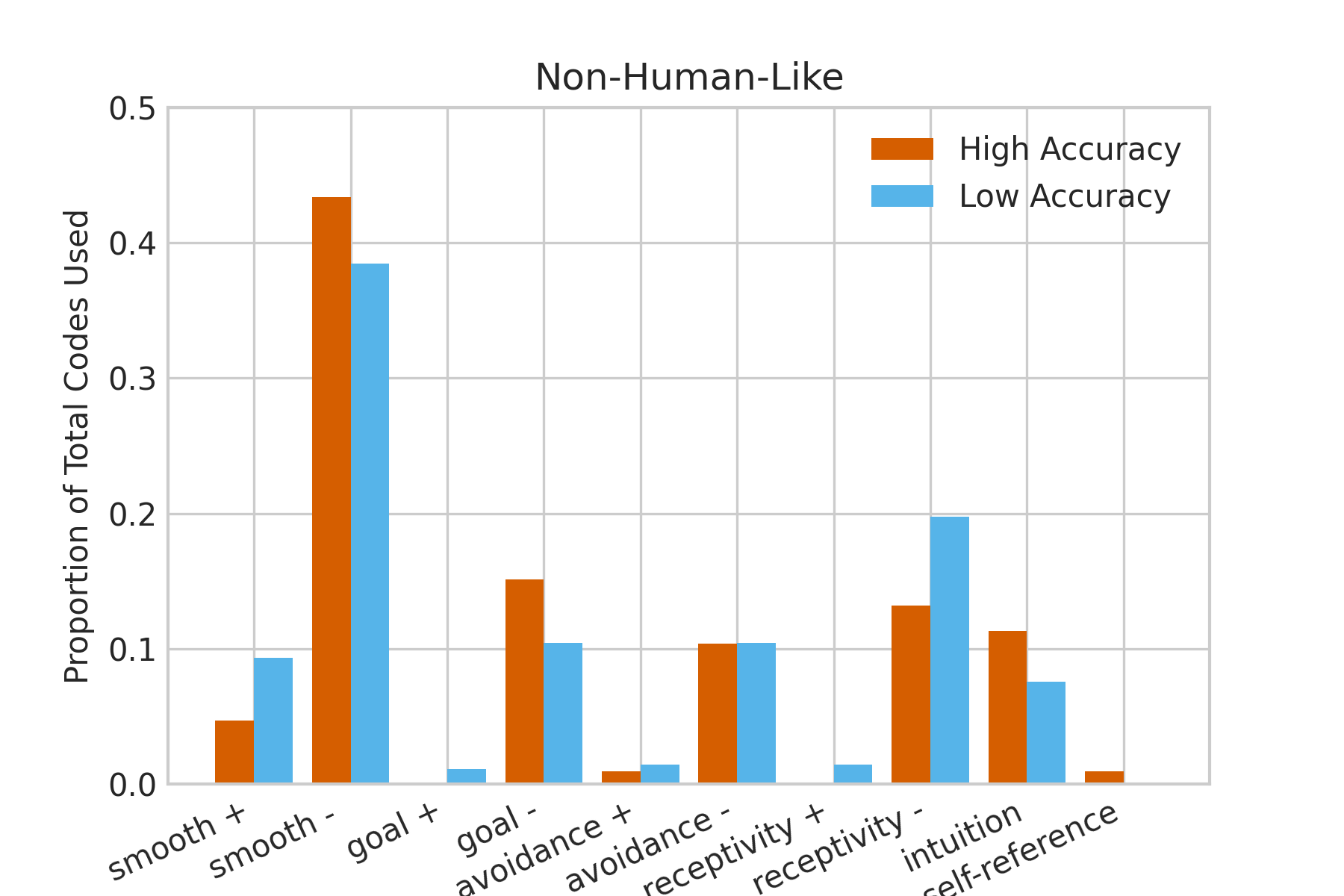}
        \caption{\textbf{Codes used to describe non-human-like behavior by low- and high-accuracy judges.} We compare the proportion of codes used to describe non-human-like behavior by human judges in their assessment of human likeness.
        }
        \label{fig:hntt-nhl-acc}
        \Description{Two bar plots showing the proportion of codes used to describe human-like and non-human-like behavior in two settings: when the participant exhibited high accuracy and when the participant exhibited low accuracy.}
    \end{subfigure}
    \hfill
    \caption{\textbf{Codes used to describe human-like and non-human-like behavior, further decomposed by high- and low-accuracy judges}. We compare the proportions of codes that are used to describe human-like behavior (left) and non-human-like behavior (right). We further decompose these codes by high- and low-accuracy judges to determine whether individuals who are more accurate rely on different features to rationalize their decisions. Interestingly, we see that the judges rely on similar characteristics to different degrees.}
    \Description{Codes used to describe human-like and non-human-like behavior, decomposed by high- and low-accuracy judges. The plots show that the judges rely on similar characteristics but to different degrees.}
    \label{fig:hntt-humanlikeness-accuracy}
\end{figure*}

\textbf{Human judges that more accurately assess human likeness exhibit different beliefs about characteristics than human judges that less accurately assess human likeness, despite relying on the high-level characteristics to similar degrees.}
We divide the participants into two groups: high-accuracy (greater than $80\%$ of responses indicating the more human-like agents aligned with the human-generated video) and low accuracy (less than or equal to $80\%$ of responses indicating the more human-like agents aligned with the human-generated video).
We examine which codes are more frequently used to describe human-likeness by the participants in each group. 
Figure \ref{fig:hntt-humanlikeness-accuracy} shows this decomposition. 
Although high- and low-accuracy judges generally rely on similar characteristics, they do so to different degrees. 
\smnote{For example, both types of judges refer to the high-level code of smoothness of movement in $40\%$ of their codes when describing human-like behavior. 
Similarly, they both refer to the high-level code of smoothness of movement in around $49\%$ of their codes when describing behavior that they do not perceive as human like.
These results indicate that there is no difference in their tendency to rely on this characteristic to explain behavior.
However, high-accuracy judges more commonly describe smooth motion when describing human-like behavior.
In contrast, low-accuracy judges more often mention smoothness as a characteristic of behavior that is not human-like.
}
This result further supports the idea that people's beliefs about AI capabilities may inform their assessments. 
In our case, the low-accuracy participants seem to share a similar belief that an AI agent is more capable than a person (by producing more smooth or ``perfect'' navigation).

Low-accuracy judges more often describe human-like agents as exhibiting more receptivity and responsiveness to their surroundings.
A similar pattern emerges with less receptivity to justify non-human-likeness. 
This result indicates that low-accuracy participants may incorrectly attribute behaviors to interacting with the environment.
As an example, a human judge that incorrectly identified Video A as being more human-like claims, 
\begin{quote}
Video A takes the more obvious route to the finish while B takes the longest possible one. A human generally would take the easiest route.
\end{quote}
Interestingly, both high- and low-accuracy judges utilize intuition and self-reference to a similar level of frequency when assessing human-like behavior.
In combination with the previous results showing that assessments of human likeness are influenced by stereotyped beliefs about AI capabilities, this finding suggests that some participants have better intuition because it aligns with the actual capabilities of AI agents. 
\section{Discussion and Future Directions}
Although conducted in a limited scope, our findings should assist with future work on designing and evaluating human-like agents.

\subsection{Limitations}
Our study specifically evaluates the human-likeness of third-person perspective point-to-point navigation behavior in agents. 
Although this type of navigation is present in many settings, like pedestrian navigation in driving simulator~\cite{walch2017evaluating} there are many other forms of navigation that exist in both real-world and virtual environments. 
Each of these types presents unique challenges and requires different strategies for designing human-like behavior. 
Although our study does not address all types of navigation, it provides a valuable starting point for evaluating the human-likeness of agents in one specific type of navigation. 
The codes that we identify are general enough to provide a starting point for researchers to analyze different forms of navigation.
For instance, collision avoidance is a general characteristic that is persistent in many domains featuring diverse types of navigation, like driving and running. 
Future work should consider expanding these evaluations to provide a more comprehensive understanding of how to design agents that behave in a more human-like manner.

Additionally, the analysis of the free-form responses revealed that there were different interpretations of the human likeness question. 
Some judges related the movement directly to human navigation in the real world. 
One judge said,
\begin{quote}
In real life a human would almost certainly not jump down as far as the character in video A did without severely hurting themselves.
\end{quote}
\noindent However, others related the movement to how human players would \textit{control} an agent in a video game. 
Another judge mentioned,
\begin{quote}
... in Video A, the player bumps into a wall briefly before readjusting. This is something humans do when they get distracted and look away for a moment.
\end{quote}

To investigate this disagreement, we annotated the agreement sample with which interpretation of the question the subject answered: real-world human navigation, video game navigation, and unclear.
The two annotators had a high agreement for this annotation (Cohen's kappa: $\kappa = 0.94$).
It was largely unclear which question the subjects were answering ($40$ out of $53$ responses). 
However $11$ responses referred to \textit{video-game} navigation, while only $2$ responses were clearly about real-world human navigation. 
We suspect that including the video game familiarity questions primed subjects to believe that the question was about video-game-specific navigation, rather than general human-like navigation.
In future studies, we recommend that the study designers clarify which question is asked of participants by including an additional question that asks the participants the other interpretation of the question to provide an obvious contrast or describing the situation in which they would like the participants to envision themselves.


\subsection{Designing and Evaluating Human-Like AI}
Our study revealed that only the agent designed to display more human-like behavior passed our test of human likeness, highlighting the importance of explicitly incorporating these objectives when designing agents. However, determining what exactly constitutes human likeness requires careful consideration from designers. 
This assertion is further supported by the different interpretations of the human likeness question by the human judges.
One interpretation of human likeness is acting as if the agent is controlled by a person, while the other refers to exhibiting more realistic behaviors. 
Both perspectives can be useful in different contexts.  

When designers seek to automate parts of the development process, such as playtesting, it is more important to create agents that appear to be human-controlled. 
In automated playtesting of games~\cite{gudmundsson2018human,sestini2022automated}, AI agents that act like real users would enable video game designers to expedite the iterative development process while also alleviating the burden of game players to extensively evaluate new content. 
Users could provide feedback only after obvious bugs, like those related to movement, have been corrected, which may enhance their enjoyment of the feedback process.
In \textit{shared autonomy}~\cite{aigner1997human}, developing agents that behave like that user would enable a more seamless integration of semi-autonomous control with user inputs. 
For example, we observed that the judges called out strafing as an example of what a human would do in a video game. 
Strafing is a tactical, sideways maneuver that would not be performed by a person navigating in the real world.
Incorporating these game-specific movements would likely increase the perception of the agent being human-controlled, especially by expert players.
\ptthree{The creation of such agents would enable players who experience disruptions, like network issues, to still play cloud games~\cite{disruption2019}.
When the system detects a disruption, it can take control and begin emulating human-like behavior. 
When the user can take back control, they can do so seamlessly.}
This can also be included as an option for players who desire in-game assistance for other reasons, such as mobility issues.
Conversely, when the objective is immersion, producing more realistic navigation is essential. 

In our study, we focused on producing more realistic navigation.
To that end, we identified a set of high-level characteristics, such as smoothness of movement, that the judges relied on to assess human likeness. 
As a result, game AI designers can first focus on adjusting these characteristics. 
\ptone{As we demonstrate with our \textit{reward-shaping} agent, these characteristics may be targeted using simple techniques} and assessed with 
an \textit{automated} Turing test~\cite{devlin2021navigation}.
After handling the most frequently mentioned characteristics, designers can then focus on more fine-grained details, such as agents not walking in puddles, to reflect more real-world navigation.

Furthermore, we employed a \textit{third-person} Turing test where participants watched videos of the agents navigating. 
Although the ability to pause, rewind, and replay the videos provided a means of interrogation, it was based solely on observation, and lacked the intervention-based approach of a typical Turing test.
Intervention-based approaches could include changing the camera perspective, adversarially interrupting the AI agent's intended path, and more. 
These forms of interaction may yield different insights. 

There are some downsides, however, to deploying a more interactive test, particularly at scale.
Recruiting a human evaluator and a human player to interact requires their simultaneous availability for real-time feedback. 
One solution is in-person studies, which can be challenging to scale and deploy. 
For instance, at the time of this study, we could not run in-person studies due to the ongoing global pandemic or distribute our proprietary game build to remote participants. 
\ptfour{Future work could take advantage of advances in game streaming, which may enable interactive remote studies with proprietary game builds. 
This solution can also incorporate previous work on simultaneous recruitment of participants~\cite{von2004labeling,bernstein2011crowds}.
However, constructing the architecture to incorporate these different technologies may require significant engineering effort.}


Importantly, previous work has demonstrated that the inclusion of more direct ways to interrogate the agent by embodying the player and agent in the same virtual space can lead to limited insight~\cite{togelius2013assessing}.
Indeed, work that included an in-game assessment of the human or bot introduced the side effect of an additional game mechanic causing some players to prioritize either gameplay or on the believability assessment~\cite{hingston2010new}.
This division of attention yields unreliable results, leading to other researchers adopting third-person variants of these assessments~\cite{arzate2018hrlb,shaker2013turing,devlin2021navigation}.
\ptone{As a result, we believe that the following pipeline could be useful for evaluating human-like agents.
Designers can initially deploy a third-person Turing test to evaluate the human likeness of specific behaviors.
The resulting characteristics can then be used to design a set of agents that exhibit different behavior that depends on the most common beliefs of the participants. 
For example, agents could move more smoothly if the participant believes smooth movement to be a feature of human likeness.
Players could then choose the characters that they want to interact with in the game, which would enable them to tailor the game to their own subjective experience and enjoyment.
This approach may offer more reliable insights into the effectiveness of the agent's design without sacrificing the integrity of the assessment process.
It could also empower game players by enabling them to exert control over their experience.}

\subsection{Toward More General Human-Like Agents}
Although the specific agent created for our study may not generalize to different games, this is a common and open challenge in the field of AI~\cite{reed2022generalist,lee2022multi}. 
Instead, we offer suggestions for using feedback from designers and players (e.g., through user research) to train human-like agents more efficiently and effectively. 

The differences in how more or less skilled human judges characterize human likeness suggests that different people have different interpretations of what constitutes human-like behavior. 
This supports the idea that the believability of NPCs in games is highly subject to the prior beliefs and expectations of the players.
This finding aligns with the fundamental principle of \textit{familiarity}~\cite{hinze2007review} that centers the real-world personal experience and knowledge of the user and implicates the importance of \textit{player-centered} design and customization~\cite{sykes2006player}.
Rather than producing monolithic human-like agents, we should strive to understand the beliefs of the player and tailor their experiences accordingly.

When moving to more complex settings, an additional difficulty is introduced.
The evaluations of human likeness become even more subjective, varying based on individual differences and cultural factors~\cite{mac2004proactive}. 
This result underscores the importance of involving diverse groups of people in the evaluation of AI agents to obtain a more comprehensive understanding of how people perceive these agents. 
In the context of games, this could look like utilizing participatory design methods~\cite{schuler1993participatory} to involve game players in the design of the AI agents themselves. 
With the consent of the players, we could use techniques in the area of learning from human feedback~\cite{ziebart2010modeling,christiano2017deep,hussein2017imitation}, which provide additional channels for people to communicate what they want from AI agents. 
With these techniques, players can provide training data to the agents in the form of preferences over paired demonstrations generated by the agent, demonstrations of the desired behavior, and more.
This can help to ensure that the AI agents are designed with the needs and preferences of diverse groups in mind.

This approach can also be used to help reduce the burden on video game designers: in complex domains, it is often challenging to specify reward signals by hand~\citep{clark2016faulty,krakovna2018specification,lehman2018surprising}. 
In part, this difficulty stems from the complexity of the desired behavior: as we have shown, human-like behavior is multi-faceted and necessitates optimizing over multiple objectives.
Furthermore, it is sometimes challenging to write down exactly what we mean when specifying a task. 
For instance, how do we construct a reward signal that captures the task of \textit{build a house in a video game in the same style as surrounding houses?}~\cite{shah2021minerl}. 
When designing a reward signal for this task, we would need to encode what counts as a house, what components are most important to emulate in the style, and which structures count as houses.
A person can quickly understand the intention of this instruction, but it is challenging to make explicit this implicit understanding.

As a result, an exciting avenue for future work involves developing more effective techniques for learning from people, evaluating user experiences of these techniques, and incorporating them into a flexible, user-friendly tool.
This tool can also help extend this work to more general game settings.
To more easily enable this line of work, assessments of human likeness could be incorporated into commonly-used game engines, like Unity~\cite{juliani2018unity}.
This tool would enable game developers to easily evaluate the human likeness of their AI agents using metrics and benchmarks that have been validated in previous research. 
Additionally, this tool could contain libraries of pretrained \textit{human-like} AI agents, which developers could use as a starting point for their own work.
For example, developers could utilize a pretrained human-like navigation agent to perform navigation but develop their own algorithm to use for different tasks. 
Using this tool could save developers time and effort by enabling them to quickly and easily create more believable and engaging agents to enhance the player experience.



\section{Conclusion}
\label{sec:conclusion}
In this work, we aimed to understand how people assess human likeness in human- and AI-generated behavior in the context of navigation in a 3D video game. 
Toward this goal, we designed and implemented a novel AI agent to produce human-like navigation behavior. 
We deployed a large-scale study of human-generated navigation behavior with three AI agents, including our novel \emph{reward-shaping} agent. 
We find that our proposed agent passes a Turing test, while the other two agents do not. 
We further investigated the justifications people provided when assessing these agents and found that people rely on similar higher-level characteristics when determining human similarity. 
In this context, we suspect that differences in the accuracy of assessing these agents are based more on fixed beliefs about the capabilities of AI systems rather than familiarity with the assessment domain of games. 
We conclude by discussing the limitations of the work, suggesting concrete design considerations for video game designers, and identifying a few critical areas for future research.

By highlighting design considerations and challenges, we hope that this paper will serve as a call for work that integrates perspectives and techniques from the HCI and AI communities. 
Building more general human-like agents requires careful design of both the agents and the evaluation protocol. 
Developing tools that can be incorporated into games and other settings enables quick iterations of these designs and the incorporation of these different techniques. 
At the highest level, we hope researchers can develop and evaluate agents that exhibit human-like behavior that improves human interaction with AI agents.

\begin{acks}
We would like to thank Evelyn Zuniga, Guy Leroy, Mikhail Jacob, Mingfei Sun, and Dave Bignell for their contributions and feedback to an earlier study in this project. 
We would also like to thank Cecily Morrison, Youngseog Chung, and Max Meijer for their helpful comments and feedback. 
We additionally thank the anonymous CHI reviewers for their detailed comments; the paper is significantly improved thanks to their suggestions.  
Co-author Fang is supported in part by NSF grant IIS-2046640 (CAREER).
\end{acks}

\bibliographystyle{ACM-Reference-Format}
\bibliography{main}
\newpage
\appendix 

\section{Details about Reinforcement Learning Agents}
\label{app:rl_agents}
In this section, we first provide information about the baseline agents used in our study.
We then provide training details for all agents.

\subsection{Baseline Agents}
\label{sec:app_architecture} 
The \emph{hybrid} and \emph{reward-shaping} agents receive an additional input of a 32x32 cropped depth buffer visual input which the \emph{symbolic} agent did not. The visuals present a third-person view of the agent in the environment. To process this additional visual channel, the \emph{hybrid} and \emph{reward-shaping} agents are equipped with a convolutional neural network which learns to extract high level visual features which are then concatenated with a representation of the symbolic inputs.

\begin{table}[t]
\centering
\begin{tabular}{ l r }
\toprule
 \textbf{Hyperparameter}  & \textbf{Value} \\ 
\midrule
Batch size & 2048  \\ 
\midrule
Dropout rate & 0.1  \\
\midrule
Learning rate & 2.5e-4  \\ 
\midrule
Optimizer & Adam~\cite{kingma2014adam}  \\ 
\midrule
Gamma & 0.99  \\
\midrule
Lambda & 0.95\\
\midrule
Clip range & 0.2  \\
\midrule
Gradient norm clipping coefficient & 0.5  \\ 
\midrule
Entropy coefficient $c_2$ & 0.0  \\ 
\midrule
Value function coefficient $c_1$ & 0.5 \\
\midrule
Minibatches per update & 4 \\ 
\midrule
Training epochs per update & 4 \\ 
\midrule
Replay buffer size & 5 x batch size \\ 
\bottomrule
\end{tabular}
\caption{\textbf{Hyperparameters for training the \emph{symbolic}, \emph{hybrid}, and \emph{\rs{}} agents.} We train all of the agents with the PPO algorithm \citep{schulman2017proximal}. For additional detail on what these hyperparameters correspond to, we encourage an interested reader to refer to the original PPO paper. We provide these hyperparameter values for reproducibility.}
\Description{Table detailing the hyperparameters for training the agents. The values are provided for reproducibility.}
\label{table:agent-hyperparam}
\end{table}

\subsection{Training Details}
\label{app:rl_training}
We provide important details about our training setup here. 

We train all three agents with PPO~\cite{schulman2017proximal}, a popular deep reinforcement learning algorithm. 
We choose PPO for training our agents for a few reasons.
This algorithm is commonly used because it is found to be empirically robust and effective in a wide range of tasks~\cite{engstrom2019implementation}.
We train each of the three agent architectures for 15 hours, the equivalent of 10 million training timesteps, on at least 3 different random seeds.
We trained all agents using Tensorflow 2.3 \citep{tensorflow2015-whitepaper} and the OpenAI Baselines PPO2 implementation \cite{baselines} with a distributed sampler.
For a full list of training hyperparameters used in all agent versions, please refer to Table \ref{table:agent-hyperparam}. We found this set to perform best on preliminary experiments. 

To effectively train agents in a complex video game setting, we use a distributed approach leveraging an in-house sample collection framework and Azure cloud resources. Training samples are collected from a scaleset of 20 low priority {GPU} virtual machines (Azure NV6), each running 3 video game instances. The samples are then sent to one training head node, a {CPU-only} Azure E32s {memory-optimized} virtual machine.

\section{Behavioral Study Details}
\label{sec:app_study}

In this section, we include additional details about our MTurk behavioral study. 

\subsection{Full Instructions}
We detail the full instructions included in the MTurk study here.
\begin{quote}
We are conducting a survey on navigation in video games for a research project. Please read the \textbf{Description} and \textbf{Requirements}, and then select the link below to complete the survey. At the end of the survey, you will receive a code to paste into the box below to receive credit for taking our survey.

\noindent \textbf{Description}:
\begin{itemize}
    \item \textbf{Overview}: The survey is anonymous and includes a required consent form, comprehension check, some background info, and 6 video sections with 3 questions each. All questions are marked *required.
    \item \textbf{Time required}: about \textbf{30 minutes}.
     \item \textbf{Compensation}: you will receive a fixed compensation of \textbf{\$6.50} for completing the task, with potential for a \textbf{\$1 bonus} for a high-quality response. For example, copy/pasting answers, or responses that are not specific to the videos on each page, will not get the bonus. 
     \item The MTurk HIT has a 1-hour duration. It will \textbf{not} allow you to submit after 1-hour has passed (\textit{remember to submit or return HITs within 1-hour so you don't time out!})
     \item If you start the task but change your mind, you may terminate your participation at any time and \textbf{return the HIT} within 1-hour, but you will \textbf{not} be paid for returned HITs or partial completions.
\end{itemize}
\textbf{Requirements}: 
\begin{itemize}
    \item You must complete all the questions.
    \item You must not have previously completed a HIT called "Navigation Turing Test (NTT)". Repeat participants are ineligible and will not be paid. 
    \item You cannot participate from tablets or mobile phones.
\end{itemize}
\textbf{Make sure to leave this window open as you complete the survey.} When you are finished, you will return to this page to paste the code into the box.
\end{quote}

\subsection{Data Collected}
In addition to the consent, familiarity, and HNTT questions, we collected the following data: timing data broken down by page and the order in which the trials were presented to each participant. 

\section{Coding Details}
\label{app:coding}
\begin{figure*}
    \centering
    \begin{subfigure}[b]{0.49\textwidth}
        \includegraphics[width=1.0\textwidth]{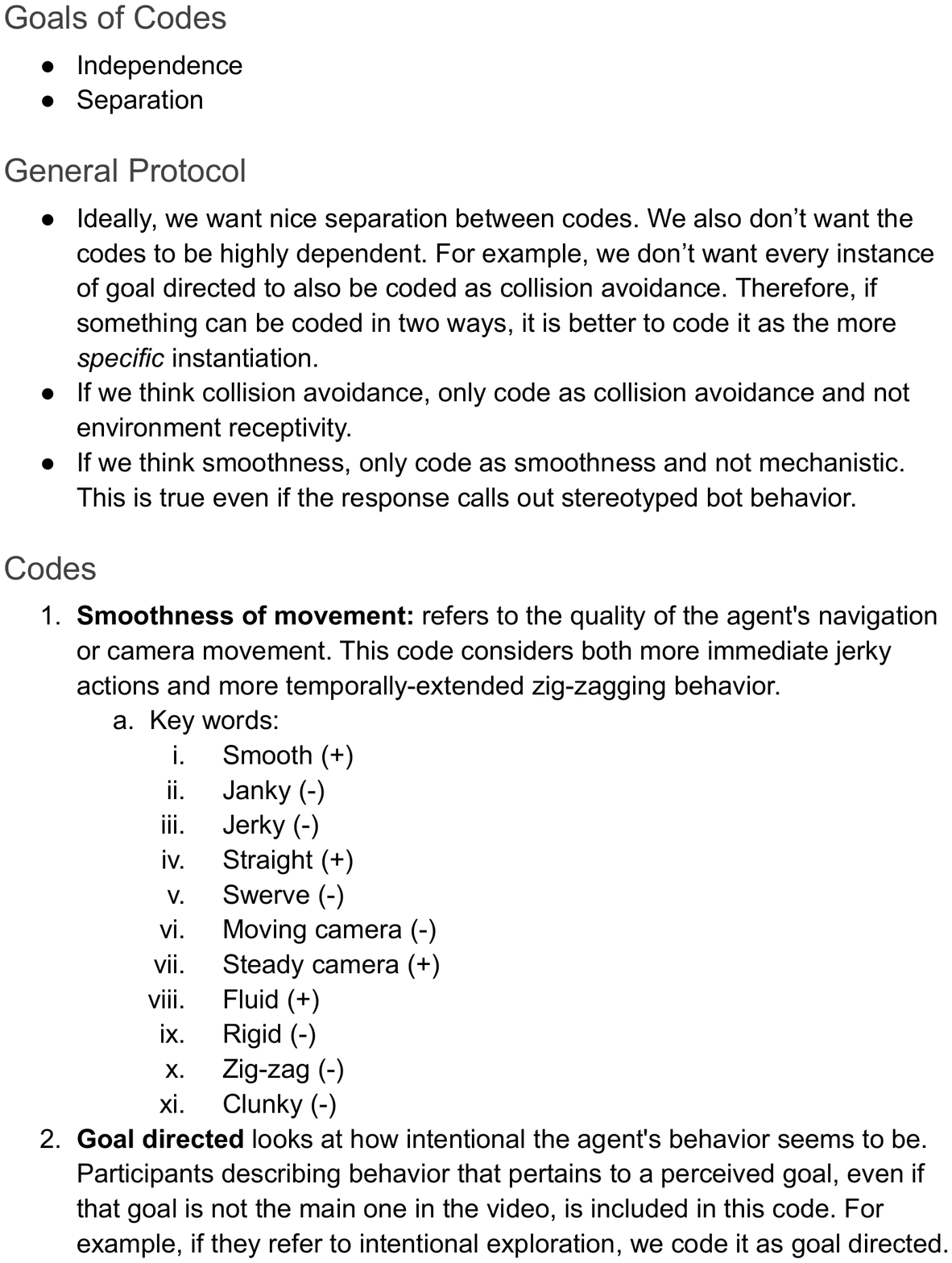}
        \Description{First page of the manual coding guideline.}
    \end{subfigure}
    \hfill
    \begin{subfigure}[b]{0.49\textwidth}
        \includegraphics[width=1.0\textwidth]{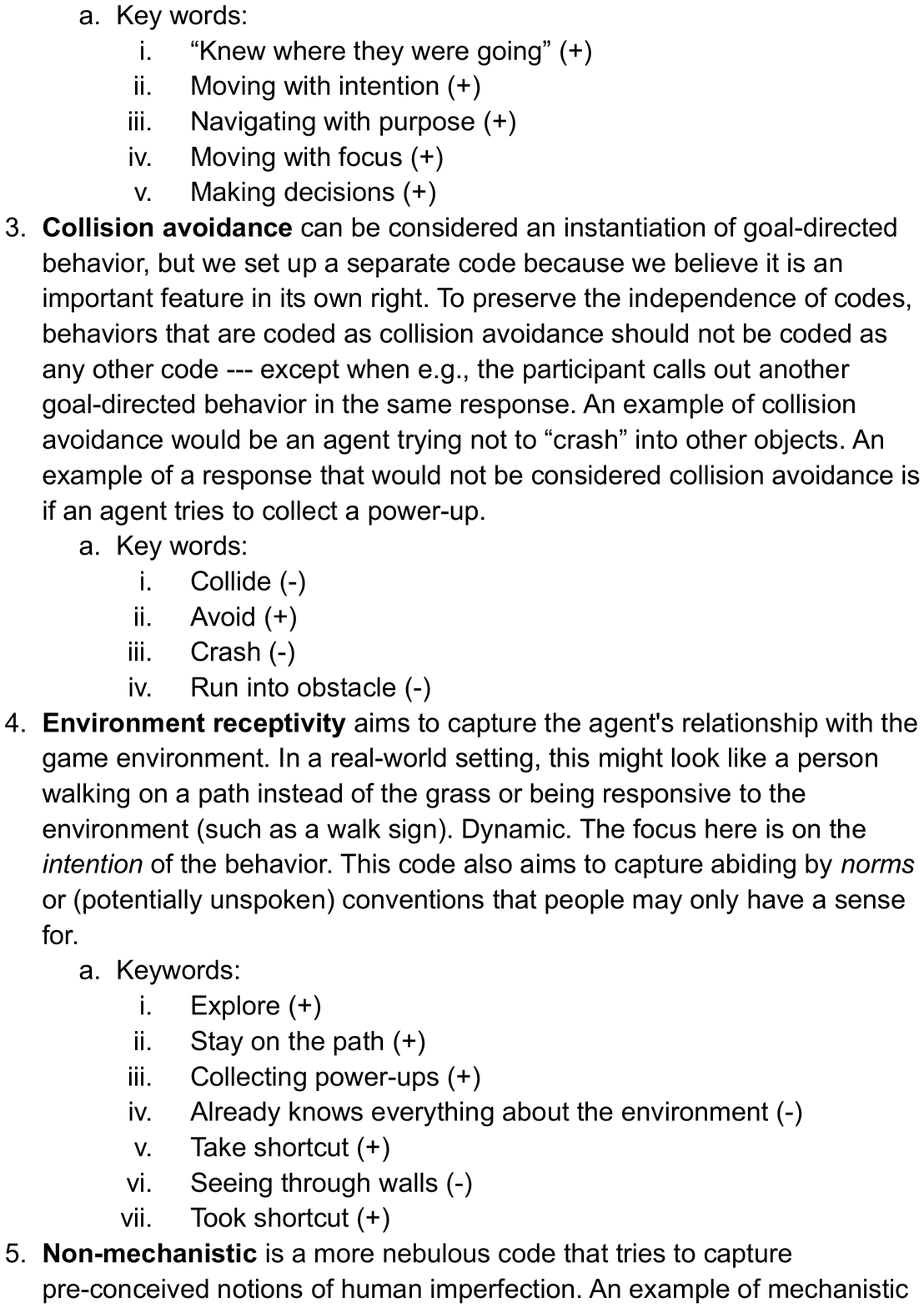}
        \Description{Second page of the manual coding guideline.}
    \end{subfigure}
    \hfill
    \begin{subfigure}[b]{0.49\textwidth}
        \includegraphics[width=1.0\textwidth]{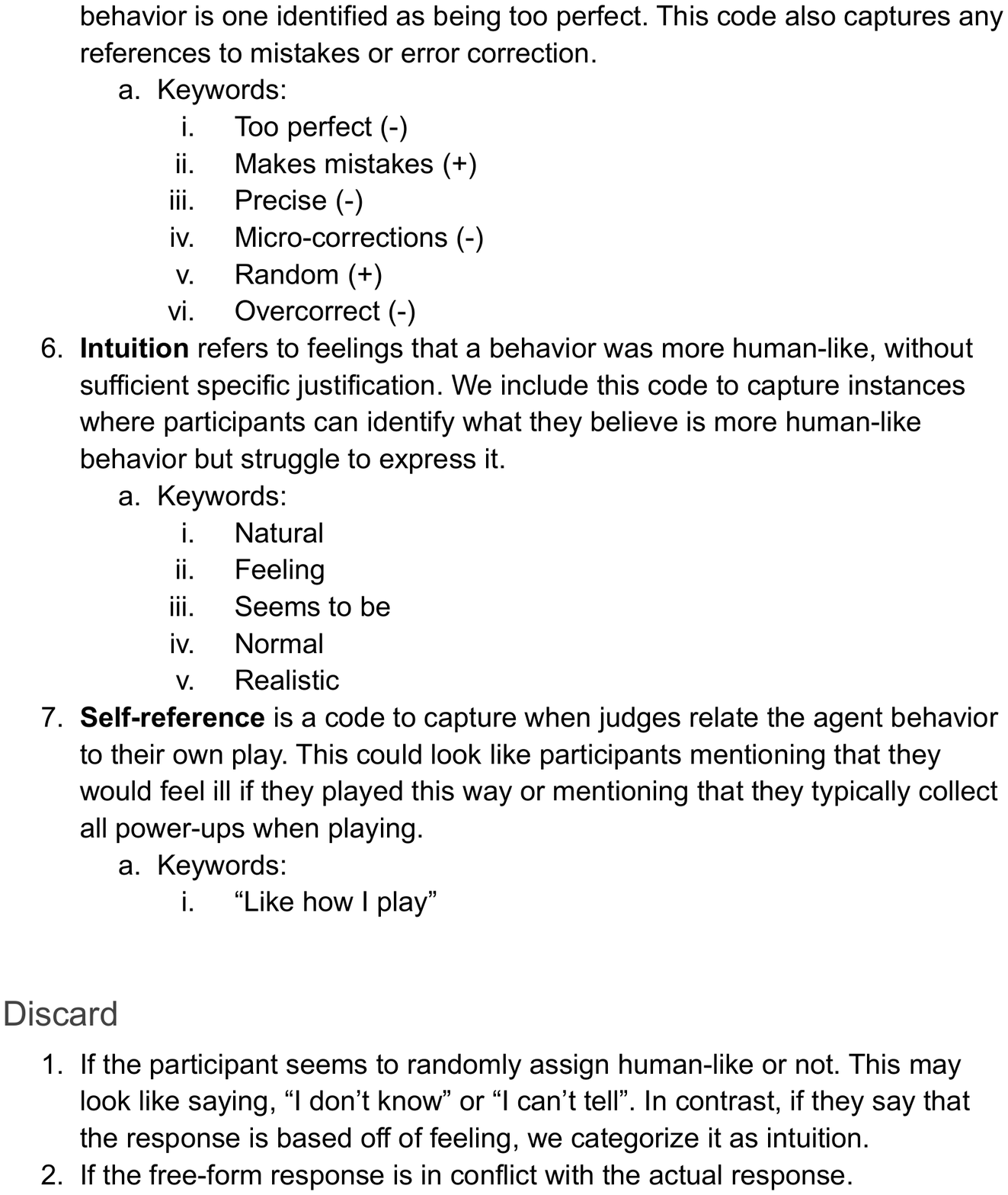}
        \Description{Third page of the manual coding guideline.}
    \end{subfigure}
    \caption{\textbf{Coding guide used by the annotators}. The guide includes a description of the codes, the general protocol, and the discard guidelines.}
    \Description{Manual coding guidelines used by the annotators. The guide includes a description of the codes, the general protocol, and the discard guidelines.}
    \label{fig:coding_guide}
\end{figure*}

We include the coding guide agreed upon and used by both annotators when annotating their responses. 
Figure~\ref{fig:coding_guide} shows a screenshot of the guide.
\end{document}